\pdfoutput=1
\documentclass[a4paper,12pt]{article}
\usepackage{amsfonts}
\usepackage{mathrsfs}
\usepackage{amsmath}
\usepackage{amssymb}
\usepackage{framed}
\usepackage[medium]{titlesec}
\usepackage{bm}
\usepackage{cite}
\usepackage[normalem]{ulem}
\usepackage{extarrows}
\usepackage{slashed}
\usepackage{booktabs}
\usepackage{isodateo}
\usepackage{graphicx}
\usepackage{xcolor}
\usepackage[bookmarksnumbered=true,bookmarksopen=true]{hyperref}
 \hypersetup{colorlinks,
             linkcolor=[rgb]{0,0.3,0.6}, 
             citecolor=[rgb]{0,0.3,0.6}, 
             urlcolor=[rgb]{0,0.3,0.6}}
\usepackage[hmargin=.7in,vmargin=1.1in]{geometry}
\usepackage{indentfirst}

\newcommand{\FR}[2]{\displaystyle\frac{\,{#1}\,}{#2}}

\newcommand{\n}{\nonumber}

\graphicspath{{fig/}}
\usepackage{enumitem}

\def\bge{\begin{equation}}
\def\ede{\end{equation}}
\def\bga{\begin{aligned}}
\def\eda{\end{aligned}}
\def\bgp{\begin{pmatrix}}
\def\edp{\end{pmatrix}}
\def\bgs{\begin{subequations}}
\def\eds{\end{subequations}}
\newcommand{\order}[1]{\mathcal{O}({#1})}
\def\di{{\mathrm{d}}}
\def\D{{\mathrm{D}}}

\def\mb{\mathbf}
\def\ms{\mathscr}

\def\pd{\partial}
\def\ld{{\mathscr{L}}}

\def\la{\langle}\def\ra{\rangle}

\def\to{\rightarrow}

\def\ii{\mathrm{i}}

\def\al{\alpha}
\def\be{\beta}

\def\de{\delta}
\def\ep{\epsilon}

\def\lam{\lambda}

\def\si{\sigma}

\newcommand{\ob}[1]{\mkern 2mu \overline{\mkern -2mu #1 \mkern -2mu}\mkern 2mu}
\newcommand{\wt}[1]{\mkern 2mu \widetilde{\mkern -2mu #1 \mkern -2mu}\mkern 2mu}

\newcommand{\beq}{\begin{equation}}
\newcommand{\eeq}{\end{equation}}
\newcommand{\blg}{\begin{align}}
\newcommand{\elg}{\end{align}}
\newcommand{\bit}{\begin{itemize}}
\newcommand{\eit}{\end{itemize}}
\newcommand{\ben}{\begin{enumerate}}
\newcommand{\een}{\end{enumerate}}

\renewcommand{\eqref}[1]{Eq.~(\ref{eq:#1})}

\newcommand{\figref}[1]{Fig.~\ref{fig:#1}}

\renewcommand{\t}{\tilde}
\newcommand{\f}{\frac}

\renewcommand{\a}{\mathsf{a}}
\renewcommand{\b}{\mathsf{b}}

\begin{document}

\title{\Large\textbf{Precision Calculation of Inflation Correlators at One Loop}}
\author{Lian-Tao Wang$^{a}$\footnote{Email: liantaow@uchicago.edu},~~~~~Zhong-Zhi Xianyu$^{b}$\footnote{Email: zxianyu@tsinghua.edu.cn},~~~~~Yi-Ming Zhong$^{c}$\footnote{Email: ymzhong@kicp.uchicago.edu}\\[2mm]
\normalsize{\emph{$^{a}$~Department of Physics, University of Chicago, Chicago, IL 60637, USA}}\\
\normalsize{\emph{$^{b}$~Department of Physics, Tsinghua University, Beijing 100084, China}}\\
\normalsize{\emph{$^{c}$~Kavli Institute for Cosmological Physics, University of Chicago, Chicago, IL 60637, USA}}
}

\date{ }
\maketitle

\begin{abstract}
  We initiate a systematic study of precision calculation of the inflation correlators at the 1-loop level, starting in this paper with bosonic 1-loop bispectrum with chemical-potential enhancement. Such 1-loop processes could lead to important cosmological collider observables but are notoriously difficult to compute due to the lack of symmetries. We attack the problem from a direct numerical approach based on the real-time Schwinger-Keldysh formalism and show full numerical results for arbitrary kinematics containing both the oscillatory ``signals'' and the ``backgrounds''.  Our results show that, while the non-oscillatory part can be one to two orders of magnitude larger, the oscillatory signal can be  separated out by applying appropriate high-pass filters. We have also compared the result with analytic estimates typically adopted in the literature. While the amplitude is comparable, there is a non-negligible deviation in the frequency of the oscillatory part away from the extreme squeezed limit. 
\end{abstract}

\section{Introduction}

Cosmic inflation is currently the leading paradigm explaining the origin of the large-scale inhomogeneity and anisotropy of our universe. The inflation is believed to take place at a very high energy scale, up to $10^{14}$ GeV in terms of the Hubble parameter $H$. At such high energies, quantum fields, including the spacetime itself, experience strong quantum fluctuations, and these fluctuations can imprint the large-scale inhomogeneity by coupling to the spacetime curvature perturbation. The $n$-point correlation of the curvature perturbation, or the primordial non-Gaussianity as it is often called, can then record the very high energy dynamics that happen during inflation \cite{Meerburg:2019qqi}. 

In particular, it has been suggested recently that the soft limits of the $n$-point correlators can be the discovery channels for heavy particles and new interactions with masses up to $\order{H}$. These studies are based on earlier works on primordial non-Gaussianities and are dubbed ``cosmological collider physics.'' \cite{Chen:2009we,Chen:2009zp,Chen:2012ge,Pi:2012gf,Gong:2013sma,Arkani-Hamed:2015bza} The general idea is that a heavy particle with $m\sim H$ can be created from the vacuum quantum fluctuation during inflation and its physical momentum then quickly redshifts to essentially zero. Being a non-relativistic state, its wave function would oscillate with a fixed physical frequency $m$, and this oscillation can interfere with the mode function of the curvature perturbation $\zeta$, producing a characteristic oscillatory signal in various soft limits of $\zeta$-correlators, including the ``squeezed limit'' ($k_3\ll k_1\simeq k_2$ where $k_i = |\mb k_i|$) of the 3-point function (bispectrum), and the ``collapsed limit'' ($|\mb k_1+\mb k_2|\ll k_i$, ~$i=1,\cdots,4$) of the 4-point function (trispectrum). Later on, this program is generalized to include new mechanisms of generating primordial fluctuations and new ways of producing heavy particles during inflation, alleviating several constraints in the vanilla slow-roll inflations and also producing particles with masses much higher than $H$ \cite{Chen:2018xck,Wang:2019gbi,Wang:2020ioa,Bodas:2020yho}. At present we have already quite a few particle physics models that are capable of generating visibly large cosmological collider (CC) signals \cite{Chen:2016nrs,Lee:2016vti,Chen:2016uwp,Chen:2016hrz,An:2017hlx,Kumar:2017ecc,Chen:2017ryl,Chen:2018xck,Wu:2018lmx,Li:2019ves,Lu:2019tjj,Liu:2019fag,Hook:2019zxa,Hook:2019vcn,Kumar:2019ebj,Alexander:2019vtb,Wang:2019gbi,Wang:2019gok,Wang:2020uic,Li:2020xwr,Wang:2020ioa,Fan:2020xgh,Aoki:2020zbj,Bodas:2020yho,Maru:2021ezc}. These signals could be searched for in the large-scale structure surveys in the near future or more futuristic 21 cm tomography from the dark ages \cite{Meerburg:2016zdz,MoradinezhadDizgah:2018ssw,Kogai:2020vzz}.

Considerations from particle physics model building show that the 1-loop process could be important for CC signals. For example, a promising class of signals comes from the chemical-potential-enhanced particles with nonzero spins, including both fermions and gauge bosons \cite{Wang:2019gbi,Wang:2020ioa}. For this class of models, the chemical potential is provided by the rolling inflaton via the axion-type dim-5 coupling to the fermions or gauge bosons \cite{Barnaby:2010vf,Barnaby:2011vw,Adshead:2013qp,Adshead:2015kza,Peloso:2016gqs,Chen:2018xck}. Such enhanced states are always in transverse polarizations and thus can enter the 3-point function only through loops. Analytical estimates showed that such loop correlators can be quite large and give rise to large CC signals \cite{Wang:2019gbi,Wang:2020ioa}. Therefore at least in this type of models, the loop process is the leading contribution to the signal while the tree-level processes are absent. It is therefore important to have a controlled and reliable procedure for computing these loop correlators. 

It turns out that the computation of cosmic correlators is highly nontrivial and challenging. There is a well-developed in-in formalism that essentially reduces the cosmic correlators to Feynman diagrams, but with several complications compared with their flat-space counterparts \cite{Weinberg:2005vy,Chen:2017ryl}. First, due to a lack of symmetry, space and time cannot be dealt with at the same footing, which renders the usual 4-momentum representation unusable. Second, the computation of correlation functions involves complicated time ordering, the complication of which grows fast with the number of interaction vertices. Third, the mode functions of particles in inflation involve products of special functions that are rather intractable analytically. Essentially due to these complications, the computation of cosmic correlators is still in a primitive stage. Compared to the highly developed and  almost industrialized Feynman diagram computation in flat space up to rather high loop orders, a systematic computation of cosmic correlators of simplest processes at tree level are only achieved recently, while the full 1-loop results are known only for a handful of simple diagrams.

The recent progress \cite{Arkani-Hamed:2018kmz,Baumann:2019oyu,Baumann:2020dch} in the analytical computation of tree-level diagrams is essentially achieved by fully exploiting the symmetry of the inflation background, which can sometimes be approximated by de Sitter (dS) space. Due to the enlarged symmetry of dS relative to a general slow-roll background, it is possible to transform a flat-space correlator into the corresponding object in dS via dS time translation and boosts, a process called ``cosmological bootstrap''. This technique is very useful for the process that respects the full dS symmetry. An alternative approach based on AdS techniques can also handle these dS invariant processes \cite{Sleight:2019mgd,Sleight:2019hfp}. There is another line of research aiming at understanding the analytical structure of cosmic correlators in recent years, or trying to go beyond the full dS covariance \cite{Pajer:2020wnj,Goodhew:2020hob,Sleight:2020obc,Jazayeri:2021fvk,Melville:2021lst,DiPietro:2021sjt,Sleight:2021plv}.

Compared with the tree-level progress mentioned above, much less is known about the 1-loop processes. Even worse, the phenomenologically important loop processes with chemical potential enhancement do not respect the full dS symmetry, which makes the aforementioned symmetry-based techniques not directly applicable. Our current understanding of these loop processes is thus only based on analytical approximations that might be applicable in special limits, but are not fully justified in more general cases. In view of this situation, it is desirable to have a fully controlled numerical computation for such processes.

In this paper, we attempt a full numerical implementation of such 1-loop diagrams at the 3-point level. We use the standard Schwinger-Keldysh (SK) formalism in real-time, which does not rely on the dS symmetries. We use the diagrammatic representation and follow the Feynman rules presented in \cite{Chen:2017ryl}, and then numerically carry out all the integrals. In particular, we do not expand the mode function around any particular point as was often done in previous analytical estimates. Instead, we adopt a piecewise expansion of the mode function, making sure that the mode function being used agrees with the full result up to controllable numerical errors.

As far as we know this is the first full numerical computation of 1-loop cosmic correlators that are relevant to cosmological collider physics. The signals from these processes have been studied analytically with several approximations made, some of which are valid only in the squeezed limit of the bispectrum and are applicable only to the oscillatory part of the process \cite{Wang:2019gbi,Wang:2020ioa}. The numerical approach we adopt here is mostly free of those approximations and thus is valid for arbitrary configurations. The numerical result we obtain can also cover both the ``background'' and the ``signal'' part of the bispectrum, and thus is useful for generating templates when confronting the theoretical predictions with data. The ``signal'' part obtained in this approach is also a useful check of the analytical estimates used in previous works.

As the first step of this numerical program, we consider bosonic processes in this paper, including the loop diagrams mediated by massive scalars and gauge bosons. We include in particular the axion-type dim-5 couplings $\phi F\wt F$ to introduce a chemical potential enhancement. There are subtleties peculiar to the numerical calculation which we will explain in full detail. Our results show that the overall amplitude can be one or two orders of magnitude larger than the oscillatory signal.  At the same time,  when the oscillation frequency is not too small, the oscillatory part of the signal can be filtered out from the full result by simple filtering techniques. We show that the oscillatory signals have overall amplitudes and scaling behavior consistent with analytical estimates. At the same time, there is a non-negligible deviation in the frequency of the oscillatory part for $k_1/k_3 < 20$.

The rest of the paper is organized as follows. In Sec.\ \ref{sec_review} we briefly review the cosmological collider physics and its observables. We first introduce the signal from a purely phenomenological point of view and then introduce the diagrammatic formalism for the actual computation. In Sec.\ \ref{sec_model} we introduce particle physics models that can give rise to CC signals at 1-loop order, including the case of a massive scalar loop and a massive gauge boson loop, both with or without chemical potential enhancement. Then in Sec.\ \ref{sec_calculation} we explain in detail the numerical calculation of these 1-loop correlators and present our results. Further discussions are collected in Sec.\ \ref{sec_conclusions}. We crosscheck some key aspects of numerical implementations using tree-level correlators in App.\ \ref{sec:crosscheckA}, discuss the behavior of the loop integrand at the large loop momentum limit in App.\ \ref{sec:large-mometum}, and crosscheck the filtering methods we adopt to separate the oscillatory and non-oscillatory parts of the signal in App.\ \ref{sec:crosscheckB}.

\section{Brief Review of Cosmological Collider Observables}
\label{sec_review}

In this section, we review the basics of the CC observables. The program of cosmological collider physics focuses on the inflaton-spectator interaction, aiming to extract the information of those spectator fields from the inflaton correlators. This is interesting in particular because the spectator fields could have too large a mass to be reached by any terrestrial experiments, but are not much heavier than the Hubble scale of the inflation so that they can be effectively produced during inflation.

\subsection{Observables}

The CC signals appear in various soft limits of $n$-point cosmic correlators. Among them, the squeezed limit of the 3-point correlator is the simplest and therefore will be our main focus. By the translation and rotation symmetries of 3-space, the 3-point correlator is a function of the triangle formed by the three external 3-momenta $\mb{k}_i~(i=1,2,3)$ and it depends on $\mb{k}_i$ only through their magnitudes, $k_i\equiv|\mb k_i|$. The squeezed limit then refers to the limit when one of $k_i$ is much smaller than the other two, e.g., $k_3\ll k_1\simeq k_2$.

For the 3-point correlator, it is conventional to define a dimensionless shape function $\mathcal{S}(k_1,k_2,k_3)$ in the following way,
\bge
  \la\zeta_{\mb k_1}\zeta_{\mb k_2}\zeta_{\mb k_3}\ra'=\FR{(2\pi)^4\mathcal{P}_\zeta^2}{k_1^2k_2^2k_3^2}\mathcal{S}(k_1,k_2,k_3),
\ede 
where a prime in $\la\cdots\ra'$ means to strip away the momentum-conserving factor $(2\pi)^3\de^{(3)}(\mb k_1+\mb k_2+\mb k_3)$, and $\mathcal{P}_\zeta\simeq 2\times 10^{-9}$ is the nearly scale-invariant power spectrum, which measures (the  square of) the size of the curvature perturbation $\zeta$. In the single-field slow-roll models, $\zeta$ is related to the inflaton fluctuation $\varphi$ via $\zeta_{\mb k}\simeq -(H/\dot\phi_0)\varphi_{\mb k}$, where $\dot \phi_0$ is the rolling speed of the inflaton background. 

The approximate scale invariance dictates that the shape function $\mathcal{S}(k_1,k_2,k_3)$ depends only on $k$-ratios up to slow-roll corrections so that it is really a function of ``shape'' rather than the ``size'' of the momentum triangle. Consequently, there are only two independent variables that the shape function can depend on. In the squeezed limit $k_3/k_1\to 0$ and $k_1\simeq k_2$, it is convenient to choose the ratio $\varrho=k_3/k_1$ and the angle $\vartheta$ between $\mb k_1$ and $\mb k_3$ as the two independent variables, so that $\mathcal{S}=\mathcal{S}(\varrho,\vartheta)$. The $\vartheta$-dependence contains the information about the angular momentum of the particles mediating the process, including both the intrinsic spin and the extrinsic one, and this angular dependence is completely fixed by angular momentum conservation. The $\varrho$ dependence, on the other hand, contains interesting information about the mass and possibly other dynamical properties of the intermediate particles. 

In fact, in the presence of a heavy intermediate particle, the shape function is generally not analytic in the $\varrho\to 0$ limit but can contain a nonanalytic piece in the form of a noninteger power $\varrho^\al$ where $\al$ is generally a complex number. Therefore, we expect that the shape function behaves like (suppressing the angular dependence) 
\bge
\label{shapeofrho}
\boxed{~~~~\vspace{3mm}
  \lim_{\varrho\to 0}\mathcal{S}(\varrho) = 
  \underbrace{A \varrho^N\big[1+\order{\varrho}\big]}_\text{analytic} + \underbrace{B\varrho^L\big[\sin\big(\omega\log\varrho+\varphi\big)+\order{\varrho}\big]}_\text{nonanalytic}.
~~~~}
\ede
Here we have several parameters, among which $A$ and $B$ are real numbers measuring the sizes of analytic and nonanalytic pieces, respectively. In general, $A$ can be either positive or negative, while we can always choose $B$ to be positive. The exponent $N$ is an integer, while $L$ is in general a real number. When $\omega\neq 0$, the exponent $L$ is often an integer or half-integer, although exceptions exist \cite{Lu:2021wxu}. In the nonanalytic piece, we see the characteristic oscillatory dependence on $\varrho$, described by two parameters, the frequency $\omega$ and the phase $\varphi$. 

It has been noticed that the dimensionless frequency $\omega$ can often be related to the mass of the intermediate particle. For example, the tree-level exchange of a scalar particle of mass $m$ gives $\omega=\sqrt{m^2/H^2-9/4}$ when $m>3H/2$. But here we emphasize that all four parameters ($B$, $L$, $\omega$, $\varphi$) can in principle be unambiguously measured, and they can provide very useful information about the dynamics of the process. Therefore it is important to calculate all these parameters from a given model. Below we discuss the analytic and nonanalytic pieces respectively.

\paragraph{Analytic Piece.}  In general, the analytic piece $A\varrho^N$ can be thought of as a result of local (point-like) interactions. The local interaction can be  either from the self-interaction of the external modes or from ``integrating out'' the intermediate particle when it is heavy ($m\gg H$). So it corresponds to the effective field theory (EFT) limit when $m\gg H$. Therefore, in the following we shall also call it  the ``EFT piece,'' although this piece is still present even when the intermediate particle is not so heavy ($m\sim$ or $<H$) and the EFT limit does not make clear sense. 

The integer power $N$ in the EFT piece contains  information about the EFT coupling of the external modes. For instance, $N=1$ when the external modes are derivatively coupled and $N=-1$ when they are gravitationally coupled. Direct (non-derivative) coupling of external modes also gives $N=-1$. But in this case, the amplitude $A$ would have a mild (logarithmic) dependence on $\varrho$ since the direct coupling softly breaks the scale invariance in the infrared (IR). 

Let us also comment on the terminology which could be rather confusing to nonexperts. We already mentioned that the shape function is a function of the shape of triangles. In literature, a particular dependence on the shape is often simply called a ``shape.'' Often appeared shapes include the local shape and the equilateral shape. Both of them are relevant to our study of cosmological collider physics and we will comment on them below. Rather confusingly, people also occasionally talked about the shape of the momentum triangle, such as the equilateral shape $k_1=k_2=k_3$, the squeezed shape $k_1\simeq k_2\gg k_3$, the folded shape $k_1\simeq k_2+k_3$, etc. These two usages of ``shape'' are completely independent. We urge the reader to exercise caution when seeing or using this term, and we refer the reader to \cite{Chen:2010xka,Wang:2013eqj} for more extensive reviews. 

The local shape refers to a collection of similar shapes that peaks in the squeezed limit $k_1\simeq k_2\gg k_3$. A typical example of local shape is
\bge
  \mathcal{S}_\text{local}\sim\FR{k_1^2}{k_2k_3}+\text{2 perms}.
\ede 
This shape is called local because it can arise from a nonlinear field redefinition $\zeta(\mb x)\to \zeta(\mb x)+\lam \zeta^2(\mb x)$ at late universe after inflation, where $\zeta(\mb x)$ is originally a Gaussian random field and $\lam$ is a number. Therefore we stress that the name ``local'' refers to a local redefinition in the late universe rather than during inflation. In other words, this is a field redefinition local to the future boundary of inflation spacetime rather than local in the bulk. A local effect on the boundary is actually very nonlocal from the bulk perspective. 

The equilateral shape refers to a collection of shapes that peak at the equilateral triangle $k_1=k_2=k_3$. A typical equilateral shape could be
\bge
\label{equilshape}
  \mathcal{S}_\text{equil}\sim \FR{k_1k_2k_3}{(k_1+k_2+k_3)^3}.
\ede
This shape is often generated by a local interaction in the bulk, such as $(\varphi')^3$. (Here a prime denotes conformal time derivative $\di/\di\tau$; see below.)

Therefore, in our parametrization of the shape function in the squeezed limit (\ref{shapeofrho}), a local shape will have $N=-1$ while an equilateral shape has $N=+1$. It is clear that the local shape is more prominent in the squeezed limit. There is of course no one-to-one correspondence  between the interaction types and the values of $N$. For example, both $(\varphi')^3$ and $\varphi'(\pd_i\varphi)^2$  would give the equilateral shape $N=1$. But this number still gives useful information about the interaction of the external modes. 

\paragraph{Nonanalytic Piece.} Although the purpose of this paper is to calculate the full three-point correlator for arbitrary kinematics, we will nevertheless pay special attention to the nonanalytic piece which encodes the on-shell particle production. This part is the main focus of cosmological collider physics, and for this reason, we will sometimes call it the ``signal'' while calling the analytic part the ``background.''

\begin{figure}[t]
   \centering
   \includegraphics[height=0.3\textwidth]{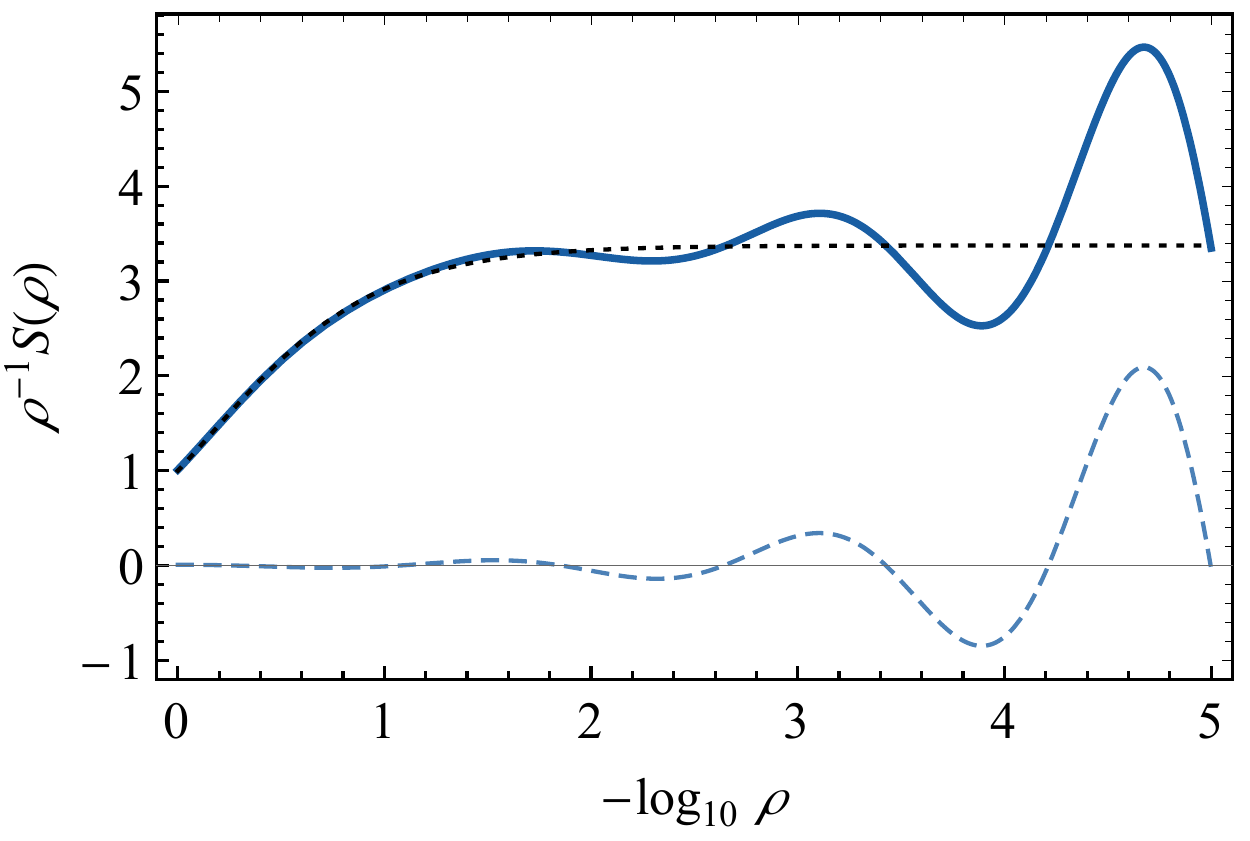} 
   \includegraphics[height=0.3\textwidth]{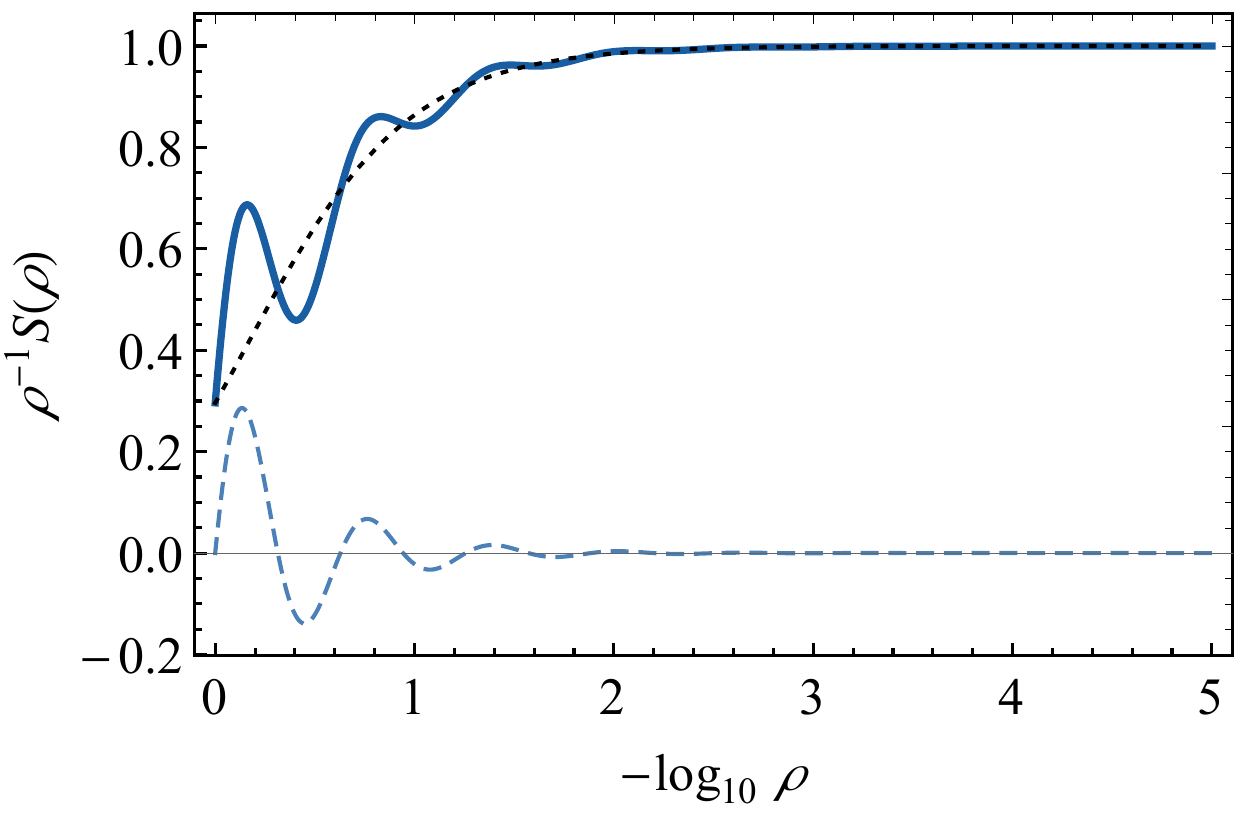}  
   \caption{The shape function $S$ of the 3-point correlator as a function of momentum ratio $\varrho=k_1/k_3$. The left and right panels show schematically the shape of the tree-level process and the 1-loop process, respectively. The solid blue, dashed blue, and dotted black curves show the full 3-point function, the signal part, and the background part, respectively. }
   \label{fig:p2}
\end{figure}

As mentioned above, the nonanalytic piece can be characterized by a 4-parameter set ($B$, $L$, $\omega$, $\varphi$). In principle, all these parameters can be calculated for specific processes and can be measured. It turns out that the scaling parameter $L$ and the oscillation frequency $\omega$ have simple parametric dependence on the model parameters such as particle's mass and spin. The overall amplitude $B$ and the phase $\varphi$, on the other hand, may have complicated dependences on these parameters. However, in some parameter regions, it is easy to estimate the leading dependence of $B$ on parameters such as mass and chemical potential. We summarize some known examples of these parameters in Table \ref{tab_signal}. This table is far from exhaustive. We use it only to illustrate how the parameters of intermediate particles (spin $s$, mass $m$, chemical potential $\mu$) are related to observables. In particular, we see that it is in general possible to tell the difference between the tree-induced and loop-induced signals, by looking at the value of $L$. This is essential because the value of $L$ measures how fast the intermediate particles are diluted by inflation. In the loop process, the particles are produced in pairs and are thus diluted faster. By examining the dilution rate of intermediate modes and how they enter the signal, we can derive, for a typical tree-level or 1-loop exchange of massive particles,\footnote{In most 1-loop processes we have $L=2$. But in the case of a Dirac fermion running in the loop and without chemical potential enhancement, the $L=2$ part of the fermion propagator happens to cancel out, and thus the leading order contribution when $\varrho\to 0$ appears with $L=3$ \cite{Chen:2018xck,Lu:2019tjj}. }
\begin{align}
\label{signalshape}
  &\lim_{\varrho\to 0}\mathcal{S}_\text{signal}^\text{(tree)}(\varrho)\sim \,\text{Re}\,\varrho^{1/2+\ii\omega},
  &&\lim_{\varrho\to 0}\mathcal{S}_\text{signal}^\text{(1-loop)}(\varrho)\sim \,\text{Re}\,\varrho^{2+\ii\omega}.
\end{align}
We plot schematically the expected shape functions from a typical tree-level process and a typical 1-loop process in Fig.\;\ref{fig:p2}. In this figure, we take the ``background part'' to be identical to the equilateral shape function (\ref{equilshape}), while the ``signal part'' takes the form of (\ref{signalshape}). The overall and relative amplitudes of the background and the signal are taken to be arbitrary in Fig.\;\ref{fig:p2}, although in given models they should be fixed and loosely related to each other. 

\begin{table}[t]
 \centering
  \caption{The parameters of nonanalytic piece in (\ref{shapeofrho}) in several representative processes mediated by particles with spin $s$, mass $m$, and chemical potential $\mu$. The value of $B$ only includes leading dependence on $m$ and $\mu$ in the large mass limit. In this table, we set $H=1$. }
  \vspace{2mm}
 \begin{tabular}{lcccc}
   \toprule[1.5pt]
    \multicolumn{1}{c}{~~~~~~}
   &\multicolumn{1}{c}{~~}
   &\multicolumn{1}{c}{$B$}
   &\multicolumn{1}{c}{~~~$L$~~~}
   &\multicolumn{1}{c}{$\omega$} \\ \hline   
    $s=0$, $m>\frac{3}{2}$, $\mu=0$ \cite{Arkani-Hamed:2015bza} & tree & $ e^{-\pi m}$ & $\frac{1}{2}$ & $ \sqrt{m^2-\frac{9}{4}}$ \\
    $s=0$, $0<m<\frac{3}{2}$, $\mu=0$ \cite{Arkani-Hamed:2015bza} & tree & -- & $\frac{1}{2}-\sqrt{\frac{9}{4}-m^2}$ & $0$ \\
    $s>0$, $m > s-\frac{1}{2} $, $\mu=0$  \cite{Lee:2016vti} & tree & $ e^{-\pi m}$ & $\frac{1}{2}$ & $ \sqrt{m^2-(s-\frac{1}{2})^2}$ \\
    $s>0$, $0<m < s-\frac{1}{2} $, $\mu=0$ \cite{Lee:2016vti} & tree & -- & \hspace{-7mm}$ \frac{1}{2}-\sqrt{(s-\frac{1}{2})^2-m^2}$ \hspace{-7mm} & $0$ \\
    $s=0$, $m>\frac{3}{2}$, $\mu=0$ \cite{Arkani-Hamed:2015bza} & 1-loop & $e^{-2\pi m}$ & $2$ & $2\sqrt{m^2-\frac{9}{4}}$ \\

    Dirac fermion, $m>0$, $\mu= 0$ \cite{Chen:2018xck} & 1-loop & $e^{-2\pi m}$ & $ 3$ & $2m$ \\
    Dirac fermion, $m>0$, $\mu> 0$ \cite{Chen:2018xck} & 1-loop & $e^{2\pi\mu-2\pi \sqrt{m^2+\mu^2}}$ & $ 2$ & $2\sqrt{m^2+\mu^2}$ \\
    $s=1$, $m>\frac{1}{2}$, $\mu\geq 0$ \cite{Wang:2020ioa} & 1-loop & $e^{2\pi\mu-2\pi m}$ & $2$ & $2\sqrt{m^2-\frac{1}{4}}$ \\
 \bottomrule[1.5pt] 
 \end{tabular}
 \label{tab_signal}
 \end{table}

\subsection{The Formalism}

We now briefly introduce the formalism that will be used for our numerical calculation in the following sections. The goal is to calculate the $n$-point correlators of the curvature perturbation $\zeta$, or equivalently, the correlators of the inflaton perturbation $\varphi$. Upon quantization, this correlator can be interpreted as the quantum expectation value of operator products $\varphi_{\mb k_1}(\tau)\cdots\varphi_{\mb k_n}(\tau)$ at the end of inflation. Throughout the paper, we work with the dS metric $\di s^2=a^2(\tau)(-\di\tau^2+\di\mb x^2)$ expressed in the conformal coordinates $(\tau,\mb x)$, and the scale factor $a(\tau)=1/(-H\tau)$. In particular, the conformal time $\tau\to (-\infty,0)$ and thus the end of inflation can be thought of as the $\tau\to 0$ limit.  We assume the standard Bunch-Davies condition for the initial state $|\text{BD}\ra$. So the $n$-point correlator can be written as $\la \varphi_{\mb k_1} \cdots\varphi_{\mb k_n} \ra$=$\la\text{BD}|\varphi_{\mb k_1}(\tau_f)\cdots\varphi_{\mb k_n}(\tau_f)|\text{BD}\ra$, where $\tau_f\to 0$ is the conformal time of the future infinity. Given a field theory model, this correlator can be calculated using the well-known Schwinger-Keldysh formalism. See \cite{Chen:2017ryl} for a review and here we only summarize the main ingredients essential to our calculation. The key observation is that the expectation value can be viewed as an ``in-in'' amplitude, and thus can be recast into a product of two ``in-out'' amplitudes with the out-state scanning over a complete basis of the Hilbert space. Each of the two in-out amplitudes is amendable to a familiar path integral representation. So the correlator can be expressed as a path integral over two sets of field variables, one goes forward in time (denoted with a `$+$' index) and the other backwards in time (denoted with a `$-$' index):
\begin{align}
\label{LPIDelta}
\la \varphi_{\mb k_1} \cdots\varphi_{\mb k_n} \ra=&~\int\mathcal{D}\varphi_{+} \mathcal{D}\varphi_{-} \,\varphi_{\mb k_1,+}(\tau_f)\cdots\varphi_{\mb k_n,+}(\tau_f)\,e^{\ii S[\varphi_+]-\ii S[\varphi_-]}  \de\big(\varphi_{+}(\tau_f)-\varphi_{-}(\tau_f)\big).
\end{align}
As above, we have two sets of fields $\varphi_\pm$, with identical action $S[\varphi_\pm]$, but with an additional minus sign in front of $S[\varphi_-]$ to account for the ``backward time''. The two sets of fields are demanded equal at the future infinity by the $\de$-function at $\tau=\tau_f$, as a consequence of summing over the out state. 

The procedure then is very similar to the usual Feynman diagram expansion of the path integral, with only a few differences which we summarize now:

First, each interaction vertex in a Feynman diagram is labeled by an SK index $\mathsf{a}=\pm$, corresponding to the two field variables $\varphi_\pm$. The minus type coupling has an additional minus sign in the vertex, coming from the minus sign in front of $S[\varphi_-]$ in the path integral. 

Second, each propagator is labeled by two SK indices at the two ends, denoted as $G_\mathsf{ab}$, and therefore we have 4 types of ``bulk'' propagators. On the other hand, if one endpoint of a propagator sits at the future boundary, then due to the identification $\varphi_+(\tau_f)=\varphi_-(\tau_f)$, the SK index at this boundary vertex does not matter. So we will have only two types of ``bulk-to-boundary'' propagators, denoted by $G_\mathsf{a}$.

Third, it is convenient to go to the 3-momentum space thanks to the 3-dim rotation and translation symmetries. However, we do not Fourier transform the time direction, so this leads to a ``mixed'' version of Feynman rules. For example, each interaction vertex is associated with a 3-momentum conservation $\de$-function, together with an integral over time $\tau$. Similarly, any loop in the diagram is associated with a 3-momentum loop integral, rather than a 4-momentum loop integral. 

The rest of the diagrammatic rules are pretty similar to the usual Feynman rules. We refer the readers to \cite{Chen:2017ryl} for more discussions on various technical details.  

\section{Bosonic 1-loop Process at the Cosmological Collider}
\label{sec_model}

The main goal of this paper is to calculate the cosmic correlator at 1-loop, with a special focus on the oscillatory signal from the loop. This section is thus devoted to a discussion of models and the related 1-loop process. This provides not only the physical motivations for our study but also the specific amplitudes that we are going to calculate in the next section.
 
 As previous studies showed (see e.g.\ \cite{Wang:2019gbi}), it is nontrivial to produce large oscillatory signals at the cosmological collider. The difficulty lies in the fact that large inflaton-matter couplings often render the matter particles too heavy to be produced, while smaller couplings reduce the signal due to vertex suppression. For many effective couplings, there is no viable parameter space in between.

 The problem of getting large signals is more acute for loop processes, not only because of the additional loop factor $1/(4\pi)^2$. In a minimal scenario, the production is supported by the inflationary expansion and is efficient for particles with mass $m\lesssim H$. When $m
\gg H$, the production is suppressed by a Boltzmann factor $e^{-\pi m/H}$. This could introduce a suppression to the CC signal. For tree-level processes, this suppression might be tolerable and a visible signal could still be produced for $m$ not much larger than $H$. But for loop processes, simple analytical estimates show that the signal is doubly suppressed, namely, by a factor of $e^{-2\pi m/H}$. This would in general make the loop signal too small to be seen. Therefore it would be desirable to consider other production mechanisms beyond the minimal scenario. 

In a well-motivated class of models, the rolling of the inflaton can produce massive particles more efficiently. See \cite{Wang:2019gbi} for discussions. In such models, the production rate is controlled by the rolling of the inflaton through a dimension-1 parameter $\mu\equiv \dot\phi_0/\Lambda$, where $\Lambda$ is a cut-off scale and the perturbativity requires $\Lambda>\dot\phi_0^{1/2}$. In typical inflation models $\dot\phi_0^{1/2}\simeq 60H$ so that the production scale $\mu$ can be as high as $60H$. For comparison, in the minimal scenario, the particle production is controlled by the cosmic expansion and is thus around the scale of Hubble $H$. 

The production of massive particles via inflaton rolling can be naturally realized for particles with non-zero spin. In such cases, axion-like couplings between the inflaton and the massive fields naturally lead to particle production at the scale of $\mu$. This includes the dim-5 couplings to a fermion $(\pd_\mu\phi)\Psi^\dag\ob\si^\mu\Psi$ and to a gauge boson $\phi F\wt F$. When evaluated with a rolling inflaton background, such operators become the number density of the corresponding matter fields weighted by the helicity of the state, with the coefficient acting as a kind of ``chemical potential.'' The size of this chemical potential is $\mu=\dot\phi_0/\Lambda$ and this explains why we have a new particle production mechanism at the scale of $\mu$.
The chemical potential enhanced particle production works only for one transverse polarization state so the corresponding spin-1/2 and spin-1 particles have to be produced in pairs. Thus their CC signals appear first at 1-loop order. (It is possible to have tree-level signals from longitudinal gauge bosons but it receives no enhancement from chemical potential.) Analytical estimates from previous studies have shown that these signals could be potentially large enough to be observed in the near future \cite{Wang:2020ioa}. However, a complete calculation of such a 1-loop process was not known. In this paper, we shall focus on the 1-loop process of the gauge boson and the result of the 1-loop fermion diagram will be presented in future work.

There is another class of models where the loop process could generate large signals. In such  models, the curvature perturbations are generated by  an additional source other than the inflaton fluctuation. Known examples of this sort include the modulated reheating scenario and the curvaton scenario \cite{Lu:2019tjj,Kumar:2019ebj}. The previously mentioned problem of double suppression is partially compensated by introducing stronger coupling between the massive particle  and the curvature perturbation. This cannot be realized in minimal slow-roll inflation because such a strong coupling would be inconsistent with the perturbativity.

\subsection{Scalar Loop}

To set the stage, we first consider the minimal case where the loop process is dS covariant. As we shall see below, the oscillatory signal in this process is in general very small. But this case is technically simpler than the more interesting cases. And also, the process is dS covariant and thus may be of theoretical interest. We are not aware of any complete analytical or numerical results for this process. The closest analytical result we have seen in the literature is a computation of 1-loop correction to the propagator in Euclidean dS completed in \cite{Marolf:2010zp}. It seems not quite straightforward to implement their result for a numerical calculation in real-time dS. So a brute-force computation directly done in real-time dS could still be useful. 

We will also consider the case with chemical potential which is observationally most interesting. However, the full dS symmetry is lost in this case, so the known analytical techniques would not apply. And it seems that our numerical approach is most appropriate in this case.

We introduce a general massive scalar particle $\si$ with bare mass $m_0$, which couples to the inflaton through the dim-6 operator $(\pd_\mu\phi)^2\si^2/\Lambda^2$. So the Lagrangian is:\footnote{ We adopt the mostly-plus metric throughout the paper.} 
\bge
\label{MinScalarLag}
  \ld=-\sqrt{-g}\bigg[\FR{1}{2}(\pd_\mu\phi)^2+V(\phi)+\FR{1}{2}(\pd_\mu\si)^2+\FR{1}{2}m_0^2\si^2-\FR{1}{\Lambda^2}(\pd_\mu\phi)^2\si^2\bigg].
\ede
Here $V(\phi)$ is the inflaton potential that generates a slow roll motion $\dot\phi_0^{1/2}\simeq 60H\simeq \text{const}$ at the background level.  This rolling generates a mass correction to $\si$ so that the $\si$ has an effective mass $m^2=m_0^2+\dot\phi_0^2/\Lambda^2$. To avoid the Boltzmann suppression (see below) we require $m\gtrsim H$. When both terms in the effective mass are positive, this implies that both terms should be at most of $\order{H}$. That is, $\Lambda^2\sim \dot\phi_0^2/H^2\simeq (60H)^2$. The resulting signal from this choice of parameter is thus very small since the 1-loop process is suppressed by $(H/\Lambda)^4$. 

For the current study, we will ignore the issue of this coupling suppression, insisting that the effective mass $m$ is of $\order{H}$. It is however possible to cook up models free from both Boltzmann suppression and coupling suppression. One example is the modulated reheating as mentioned before. Another possibility is that the dim-6 operator has a flipped sign than it is in (\ref{MinScalarLag}). Then it is possible to have both $m_0$ and $\dot\phi_0^2/\Lambda^2$ large but $m_0^2-\dot\phi_0^2/\Lambda^2$ remains $\order{H}$. This of course involves fine-tuning at the EFT level. But one can imagine that the theory contains a dense spectrum of scalar particles with masses around $\dot\phi_0/\Lambda$ and mass differences being $\order{H}$ or smaller. Then the tuning is automatically implemented. One can further introduce a quartic self-interaction for these massive scalars to bound those states with negative effective mass. We leave a concrete model building of this sort to future works. 

It is clear that the leading order contribution to the 3-point inflaton correlator from $\si$ field is at 1-loop order, 
\bge
\label{fd_scalarloop}
  \parbox{0.38\textwidth}{\includegraphics[width=0.38\textwidth]{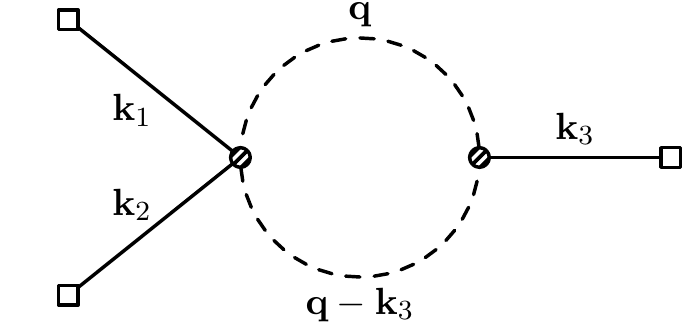}}
\ede
The diagram is ultraviolet (UV) divergent and needs to be regularized, which we will discuss below.  

In the numerical calculation in the next section, we will also include a nonzero chemical potential to the scalar field. In realistic model building, one needs additional symmetry breaking in order to introduce a chemical potential to the scalar field at the level of mode functions. For example, \cite{Chua:2018dqh} considered a possibility with a background static electric field acting as a chemical potential to the scalar field. This setup breaks rotation invariant and the resulting chemical potential is anisotropic. We are currently not aware of any scalar potential that does not break additional spacetime symmetries and works at the level of mode functions\footnote{Ref.\ \cite{Bodas:2020yho} introduces a scalar chemical potential of a different type which does not work at the level of mode functions but enhances the signal through particle injection at interaction vertices.}. So an isotropic chemical potential of the scalar field can only be viewed as a toy example.

\subsection{Gauge Boson Loop}

The second process we are going to consider is the 3-point correlator mediated by a massive gauge boson at 1-loop. The gauge boson obtains its mass from a Higgs. We find it technically simpler to couple the gauge boson to the inflaton indirectly through the Higgs. Therefore, we consider the following Lagrangian
\begin{align}
\label{gaugebosonmodel}
  \ld=\sqrt{-g}\bigg[-\FR{1}{4}F_{\mu\nu}F^{\mu\nu}-|\D_\mu\Sigma|^2-\lam|\Sigma|^4-\FR{1}{2}(\pd_\mu\phi)^2-V(\phi)-\FR{1}{\Lambda^2}(\pd_\mu\phi)^2|\Sigma|^2\bigg].
\end{align}
That is, we consider a scalar QED with $U(1)$ scalar $\Sigma=(\si+\ii\pi)/\sqrt{2}$ that couples to the inflaton $\phi$ through the dim-6 operator introduced in the previous subsection. The difference from the previous case is that the Higgs field $\Sigma$ now has a nonzero vacuum expectation value $v^2\equiv\la\si\ra^2=\dot\phi_0^2/(\lam\Lambda^2)$. Consequently, there will be a 2-point mixing between $\si$ and the inflaton fluctuation $\varphi$. This is essentially the model considered in \cite{Wang:2020ioa} except that we do not include the dim-5 operator $\phi F\wt F$ here. We will include this dim-5 operator later. { One can also include a bare mass term for $\Sigma$ which in general does not alter the qualitative picture. We neglect this term here for simplicity and refer the readers to \cite{Wang:2020ioa} for a more complete treatment.}

Here a technical remark needs to be made. As long as the oscillatory signals are the only concern, we can work in the unitary gauge. The oscillatory signals are from on-shell particle production and the corresponding loop integral is never UV divergent. More explicitly, we know that the UV divergent pieces in our models can always be subtracted by local counterterms and these local counterterms always generate analytic dependence on external momenta. Therefore, the nonanalytic momentum dependence in the signal must be free of UV divergence. 

In our current work, however, we are aiming at a complete calculation of the bispectrum, including both the smooth ``background'' and the oscillatory signal. So we need to pay attention to the UV part since the background part is UV divergent just like its flat-space counterpart. Then the calculation would be tricky in the unitary gauge. Therefore we need to consider a general $R_\xi$ gauge condition $G=\nabla_\mu A^\mu-\xi g v\si$ where $g$ is the gauge coupling and $\xi$ is a gauge parameter. After evaluating the Lagrangian in the dS background and standard quantization procedure, we find the Lagrangian for the fluctuating fields as
\begin{align}
  \ld=&~\FR{1}{2}A^\mu\square A_\mu-\FR{1}{2}a^2(g^2v^2+3H^2)A^\mu A_\mu-a^2(\pd_\mu\bar c)(\pd^\mu c)-a^4g^2v^2\bar cc\n\\
  &-\FR{1}{2}a^2\Big[(\pd_\mu\si)^2+(\pd_\mu\pi)^2+(\pd\varphi)^2\Big]-\FR{1}{2}a^4m_\si^2\si^2-\FR{1}{2}a^4g^2v^2\pi^2\n\\
  &-a^4\lam v\si(\si^2+\pi^2)-\FR{1}{4}a^4\lam(\si^2+\pi^2)^2+\FR{\dot\phi_0v}{\Lambda^2}a^3\si\varphi'\n\\
  &+a^2gA^\mu(\si\pd_\mu\pi-\pi\pd_\mu \sigma)-a^2g^2v\si A^2-\FR{1}{2}a^2g^2(\si^2+\pi^2)A^2-\FR{1}{2}a^4g^2v\si\bar cc+\cdots,
\end{align}
where $c$ is the ghost field. We have taken the Feynman-'t Hooft gauge $\xi=1$ and neglected interaction terms irrelevant to us. All scale factors have been spelled out explicitly and so the spacetime time indices are raised or lowered by Minkowski metric $\eta_{\mu\nu}$, except in the kinetic term of the gauge field, where $\square=g^{\mu\nu}\nabla_\mu\nabla_\nu$.

Then at the 1-loop level, we can consider the gauge boson signal from the following two diagrams.\footnote{ Here we only consider diagrams in which the gauge bosons couple only to the Higgs $\si$. There is another class of diagrams arising from $\phi F\wt F$ coupling. These diagrams give rise to similar oscillatory signals but different angular dependence. In \cite{Wang:2020ioa} it was shown that they are more suppressed than the ones considered here. For this reason we do not include these diagrams in the current paper.}
\bge
\label{fd_gaugeloop}
  \parbox{0.38\textwidth}{\includegraphics[width=0.38\textwidth]{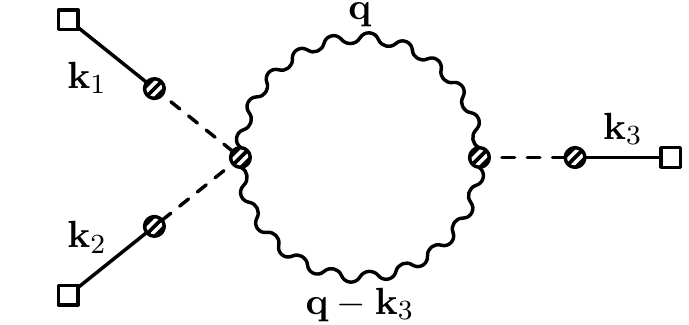}}~~~~~~
  \parbox{0.38\textwidth}{\includegraphics[width=0.38\textwidth]{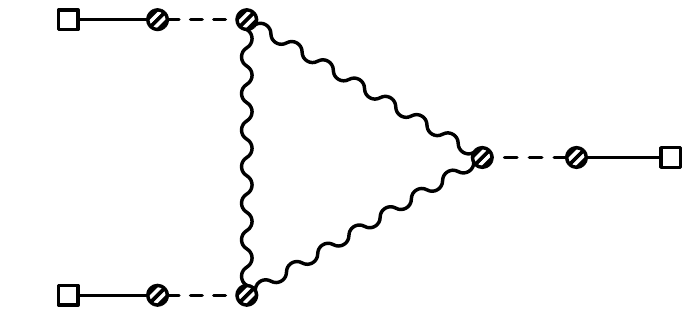}} 
\ede
Here we only show the gauge boson diagrams. There are corresponding diagrams with $\pi$ and ghost fields. We can drop some of these diagrams by considering special limits. For example, when $\lam\gg g^2$, we only need to include $\pi$-graphs. But this would be identical to a scalar loop considered in the previous subsection and nothing new could be learned. We can also consider a different limit with $\lam\ll g^2$ and $v\gg H$. In this limit, we can drop all graphs containing $\pi$, and retain diagrams with the gauge boson loop and the ghost loop.  Of course, there can be more interesting cases if the gauge boson is coupled to external line differently (i.e., through derivative couplings). We leave those possibilities for future study.

There are additional complications in the two diagrams in (\ref{fd_gaugeloop}) which are usually nonessential to the CC signals but increase the difficulty of numerical calculation. One is the presence of the dashed lines in the external legs in all diagrams. We can approximate these dashed lines by their EFT limit $1/m^2$, assuming its mass is greater than $H$. Another complication is the triangle loop in the second diagram in (\ref{fd_gaugeloop}), which requires one more layer of time integral. In this work we shall focus only on the left diagram in (\ref{fd_gaugeloop}) which we call the pinched diagram, leaving a complete evaluation of the right diagram (the triangular diagram) for future work. Incidentally, if we are only concerned with the signal part of the result, then we can approximate one of the three internal lines in the triangular diagram by its EFT limit. { Thus generated ``pinched coupling'' is a result of integrating out one of the three internal lines, and we expect that the main contribution to this integral comes from the region when the two endpoints are separated by a distance within Hubble radius. }  The resulting graph is then identical to the left (pinched) diagram in (\ref{fd_gaugeloop}).

Finally, we add the following operator into the Lagrangian (\ref{gaugebosonmodel}) which acts as a chemical potential for the helicity of gauge bosons,
\bge
  \Delta\ld=\FR{1}{\Lambda_F}\phi F\wt F.
\ede
This additional term modifies the dispersion relation of the gauge boson modes during inflation and thus can enhance the signal. The relevant diagrams are still the two in (\ref{fd_gaugeloop}); only the propagators of the loop lines need to be replaced by the one including  the chemical potential $\mu=\dot\phi_0/\Lambda_F$. We refer readers to \cite{Wang:2020ioa} for more details.

\section{Numerical Implementation and Result}
\label{sec_calculation}

In this section, we present a detailed treatment of the numerical integration. Although we focus on the pinched diagram throughout the paper, the procedures and techniques introduced here can be generalized to other 1-loop computations. We first present the expressions for the 1-loop integrals and then discuss their numerical implementations. Finally, we present the numerical results.

\subsection{One-Loop Integral} 

\begin{figure}[htbp]
   \centering
   \includegraphics[width=0.48\textwidth]{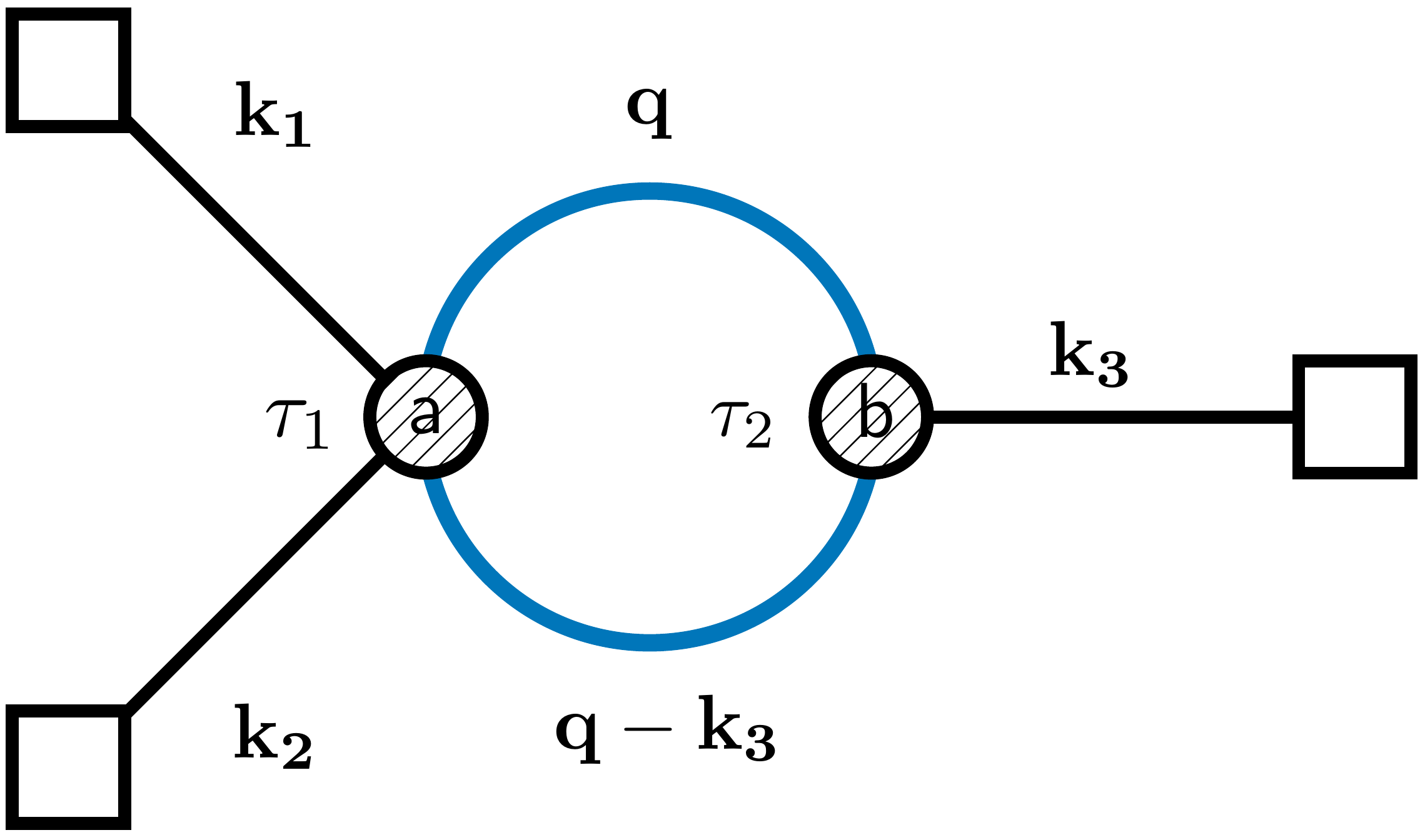} 
   \caption{The pinched diagram. $\a$ and $\b$ are SK indices. $\mb k_{1,2,3}$ are external 3-momenta. $\tau_1, \tau_2$ are conformal time variables. The black and blue lines represent the boundary-to-bulk propagators and the internal loop propagators respectively. }
   \label{fig:p3}
\end{figure}

All the diagrams from the last subsection, including the triangle diagram after pinched approximation, can be put into the following form 
\begin{align}
  \mathcal{B}(k_1,k_2,k_3)= &~\FR{1}{2}
 \sum_{\mathsf{a},\mathsf{b}=\pm}\mathsf{ab} (\ii)^2 \int\FR{\di\tau_1}{|H\tau_1|^\al} \FR{\di\tau_2}{|H\tau_2|^\be}\pd_{\tau_1} G_\mathsf{a}(k_1;\tau_1)\pd_{\tau_1} G_\mathsf{a}(k_2;\tau_1)\pd_{\tau_2}G_\mathsf{b}(k_3;\tau_2)\mathcal{I}_{\a\b}^{(1)}(k_3;\tau_1,\tau_2)\n\\
  &~+(k_1\leftrightarrow k_3) + (k_2\leftrightarrow k_3),
  \label{eq:vectormaster}
\end{align}
where the last line represents momentum permutations, { which appear because of the three possible ways to connect the loop with the three external legs}. In Fig.\;\ref{fig:p3} we show the diagram corresponding to the first line of (\ref{eq:vectormaster}).
Here we have suppressed all the coupling dependence that is nonessential to numerical computation. The related CC signal is given by $\mathcal{S}(k_1, k_2, k_3)= k_1^2 k_2^2 k_3^2 \mathcal{B} (k_1, k_2, k_3)$. As explained above, we need to assign SK indices $\mathsf{a}$ and $\mathsf{b}$ to the two vertices, respectively. The factor $|H\tau_1|^{\al}$ and $|H\tau_2|^{\be}$ in (\ref{eq:vectormaster}) come from the $\sqrt{-g}$ factor in the Lagrangian as well as additional $g_{\mu\nu}$ factors in the vertices. We use $G$ to denote the propagator of the external (inflaton) lines:
\bge
  G_\pm (k;\tau)=\FR{H^2}{2k^3}(1 \mp\ii k\tau)e^{\pm\ii k\tau}.
\ede
In all cases, we include the derivative coupling to the inflaton by using $\pd_\tau G_\mathsf{a}(k;\tau)$ for the three external legs. For the scalar model, there is in principle a contribution from spatial derivative in the form of $\pd_i G\pd^i G$ at $\tau_1$-vertex. We will drop this term for simplicity { although the contribution from $\pd_i G\pd^i G$ is of the same order as $\pd_\tau G\pd_\tau G$}. The loop integrals $\mathcal{I}_\mathsf{ab}^{(1)}$ are given by
\beq
  \mathcal{I}_{\a\b}^{(1)}(k;\tau_1,\tau_2)=\int\FR{\di^3\mb q}{(2\pi)^3}D_{\a\b}(q;\tau_1,\tau_2)D_{\b\a}(p;\tau_2,\tau_1).
\eeq
where $D_{\a\b}(k; \tau_1, \tau_2)$ are the bulk-to-bulk propagators
\bgs
\label{eq:SKprop}
\begin{align}
D_{++}(k;\tau_1,\tau_2)=&~D_>(k;\tau_1,\tau_2)\theta(\tau_1-\tau_2)+D_<(k;\tau_1,\tau_2)\theta(\tau_2-\tau_1),\\
D_{+-}(k;\tau_1,\tau_2)=&~D_<(k;\tau_1,\tau_2),\\
D_{-+}(k;\tau_1,\tau_2)=&~D_>(k;\tau_1,\tau_2),\\
D_{--}(k;\tau_1,\tau_2)=&~D_<(k;\tau_1,\tau_2)\theta(\tau_1-\tau_2)+D_>(k;\tau_1,\tau_2)\theta(\tau_2-\tau_1),
\end{align}
\eds
with $D_>(k; \tau_1, \tau_2) \equiv u(\tau_1, k) u^*(\tau_2, k)$ and $D_<(k; \tau_1, \tau_2) \equiv u^*(\tau_1, k) u(\tau_2, k)$ where $u(\tau,k)$ is the mode function for the loop particle, either scalar or gauge boson. It turns out that the four loop integrals ($\mathcal{I}_{++}^{(1)}$, $\mathcal{I}_{+-}^{(1)}$, $\mathcal{I}_{-+}^{(1)}$, and $\mathcal{I}_{--}^{(1)}$) are all related to the following unique integral by including various time ordering and complex conjugation:
\begin{align}
\label{loopint}
  \mathcal{I}_{-+}^{(1)}(k;\tau_1,\tau_2)=\int\FR{\di^3\mb q}{(2\pi)^3}D_>(q;\tau_1,\tau_2)D_>(p;\tau_1,\tau_2).
\end{align}
The detailed form of  $D_>$ will be given below. We will compute numerically the following cases:

\begin{description}
\item[Scalar loop.] As mentioned above we will include a fictitious chemical potential $\mu$ to enhance the signal. In this case, we have $\al=2$, $\be=3$ where $\al,\be$ are defined in (\ref{eq:vectormaster}), and 
\bge
  D_>(k;\tau_1,\tau_2)= \frac{e^{\pi \tilde \mu}}{2k} |H \tau_1| |H \tau_2| \text{W}_{- \ii \tilde \mu, \ii \tilde \nu} (2 \ii k \tau_1)\text{W}_{\ii \tilde \mu, \ii \tilde \nu} (-2 \ii k \tau_2),
  \label{eq:scalarprop}
\ede
where W is the Whittaker W function, $\wt\mu\equiv\mu/H $ and $\wt\nu\equiv\sqrt{(m/H)^2-9/4}$. We only consider scalars with mass $m>3H/2$ so $\wt\nu$ is always real. 

\item[Gauge boson loop without chemical potential.] For this example, we will stay in the Feynman-'t Hooft gauge ($\xi=1$). We only consider the limit $\lam\ll g^2$ and $v\gg H$, so that the Goldstone loops decouple, and we only need to include the gauge boson loop and the ghost loop. It turns out that the ghost loop and the $A^0$ loop contribute the same fictitious oscillation pattern at superhorizon scales and the two contributions cancel each other. So only the three physical degrees of freedom in the massive gauge boson contribute to the signal.  In this case, we have $\al=0$, $\be=1$. The propagator will be taken to be, 
\begin{align}
  D_{\mu\nu >}(k;\tau_1,\tau_2)={}& g_{\mu\nu} D_{>}^{\text{(spin-1)}} (k;\tau_1, \tau_2)  = \frac{g_{\mu\nu}}{2k}  \text{W}_{0, \ii \tilde \nu} (2\ii k \tau_1) \text{W}_{0, \ii \tilde \nu} (-2\ii k \tau_2) \nonumber \\
  ={}& \frac{g_{\mu\nu} \pi e^{-\pi \tilde \nu} (\tau_1 \tau_2)^{1/2}}{4}  \text{H}_{\ii \tilde \nu}^{(1)}(-k \tau_1) \text{H}^{(2)}_{-\ii \tilde \nu} (-k \tau_2),
  \label{eq:vectorprop}
  \end{align}
where $\text{H}^{(1)}$ ($\text{H}^{(2)}$) is the Hankel function of the first (second) kind. For the gauge boson of spin-1, we have $\wt\nu=\sqrt{(m/H)^2-1/4}$.
The indices of propagators should thus be contracted in the integrand of the loop integral (\ref{loopint}), namely, we replace $D_>(q;\tau_1,\tau_2)D_>(p;\tau_1,\tau_2)\to D_{\mu\nu>}(q;\tau_1,\tau_2)D_>^{\mu\nu}(p;\tau_1,\tau_2)$. Again the indices here are raised by flat Minkowski metric. The propagator being proportional to $g_{\mu\nu}$ is a consequence of choosing $\xi=1$ gauge. This naively introduces $g_{\mu\nu} g^{\mu\nu}=4$ propagating degrees when contracting the loop propagators. One of these 4 propagating degrees is nevertheless subtracted by the ghost loop. So we will further take $D_{\mu\nu>}(q;\tau_1,\tau_2)D_>^{\mu\nu}(p;\tau_1,\tau_2)\to 3D^{\text{(spin-1)}}_>(q;\tau_1,\tau_2)D^{\text{(spin-1)}}_>(p;\tau_1,\tau_2)$.

\item[Gauge boson loop with chemical potential.] In this case, we still have $\al=0$, $\be=1$. Due to the presence of a nonzero chemical potential, one transverse polarization will be exponentially enhanced relative to the other two polarizations. Therefore we can keep only this enhanced helicity state to a good approximation. The propagator in this case is
\begin{align}
  D_{\mu\nu >}(k;\tau_1,\tau_2)={}& \sum_{h} e^{(h)}_\mu (\mb k)e^{(h)*}_\nu (\mb k) D_{>}^{(h)}(k;\tau_1, \tau_2) \approx e_\mu^- (\mb k) e^{-*}_\nu (\mb k) D_{>}^{-}(k;\tau_1, \tau_2) \nonumber\\
  ={}&e_\mu^- (\mb k) e^{-*}_\nu (\mb k) \f{e^{\pi \tilde \mu}}{2k} \text{W}_{-\ii \tilde \mu, \ii \tilde \nu} (2\ii k \tau_1) \text{W}_{\ii \tilde \mu, \ii \tilde \nu} (-2\ii k \tau_2),
    \label{eq:vectorprop2}
\end{align}
where $e^{(h)} (\mb k)$ is the polarization vector for helicity state $h$.

\end{description}

In all cases above, the loop integral is divergent in the UV just like in flat space. However, the asymmetric treatment of space and time makes it difficult to isolate and extract the divergence directly from the integrand. Therefore we have to choose a regularization method that is straightforward for numerical implementation. We will adopt a Pauli-Villars regulator, by rewriting the loop integral (\ref{loopint}) as 
\begin{align}
  \mathcal{I}_{-+}^{(1)}(k;\tau_1,\tau_2)=\int\FR{\di^3\mb q}{(2\pi)^3}D_\text{reg}(q;\tau_1,\tau_2)D_\text{reg}(p;\tau_1,\tau_2),
\end{align}
where the regularized propagator $D_\text{reg}$ is the original propagator with mass $m$ subtracted by the same propagator but with a heavier mass $M$,
\bge
\label{eq:pauli}
  D_\text{reg}(q;\tau_1,\tau_2)\equiv D_m(q;\tau_1,\tau_2)-D_M(q;\tau_1,\tau_2).
\ede
Then the heavy mass $M$ serves as an effective cutoff of the loop integral. 

After the regularization, one still needs a renormalization scheme to fix the answer of the loop correction. This can in principle be done in given models. Since this part is less relevant to our numerical computation and is more model-dependent, we will only comment briefly on it here. Take the scalar loop model (\ref{MinScalarLag}) as an example, one needs to include a counterterm 
\bge
  \Delta\ld\supset \de_C (\pd_\mu\varphi)^2\varphi'
\ede
in order to cancel the regulator dependence in the loop integral. Then a renormalization scheme is needed to fix the regulator-independent part of the counterterm coefficient $\de_C$. As we will see below, the nonanalytic signal part of the correlator is free from the UV divergence and is independent of the renormalization scheme, since it is essentially from the on-shell particle production. Therefore it is at least in principle possible to extract the mass and the coupling of the intermediate particles by measuring the oscillatory signal. On the other hand, the background part of the signal receives UV contributions and its loop correction will be renormalization-dependent. The renormalization condition can thus be determined by measuring the overall amplitude of the background part of the correlator, namely the coefficient $A$ in (\ref{shapeofrho}). Then, using the measured mass, coupling, and amplitude $A$, one can fix the coefficient $\de_C$. { To tell apart the background from the signal, one needs to measure a range of external momentum configurations, namely, to measure the shape dependence. After the above subtraction is done, one can then use the loop diagram result plus the counterterm to predict the shape function for wider range of shapes. While this procedure is in line with the usual renormalization story in flat space, an important difference needs to be noted: In flat-space QFT, changing the external momenta amounts to changing the energy scale. For inflation correlators, however, the external momenta only label the comoving scales but not the physical energy scales. Therefore, changing external comoving momenta only probes the shape dependence in a scale invariant model, whereas the process always happens at a fixed physical scale (namely the scale of particle production at either the chemical potential $\mu$ or Hubble scale $H$.)

}

A complete set of renormalization conditions of course requires more careful treatment of various loop diagrams contributing to inflaton's correlators. The above simple argument nevertheless tells us that the background part of the 3-point function, in general, cannot be calculated unambiguously without a careful renormalization procedure. On the other hand, the signal part is free from UV issues. Below we will see that this is indeed the case by showing that the signal part is regulator-independent while the background part is regulator-dependent. Therefore, the background part of our calculation below should be treated only as an indication of the overall size of the regularized loop diagram. One should not view them as the direct prediction of the model and should not compare them with data directly.

\subsection{Setup and Procedure}
\label{sec_setup}
We separate the numerical evaluation of the integral (\ref{eq:vectormaster}) into two stages:
\begin{enumerate}[label=(\roman*)]
\item For each combination of SK indices  $({\a\b})$, we  generate a discrete 2-dim grid of the conformal time $(\tau_1, \tau_2)$ and compute the loop integral $\mathcal{I}_{\a\b}^{(1)}$ on each grid point (with Wick rotation as introduced below). We parallelize the evaluation of the grids on a cluster. This stage is computationally expensive, given the grids are tightly spaced. We utilize several properties to reduce the computation load. First, given the reality condition of SK diagrams, we only need to compute the grids with SK indices of $({\a\b})=(++)$ and $(+-)$. No evaluation for $({\a\b})=(--)$ or $(-+)$ is needed. Second, as explained in detail later, we only need to evaluate the $\tau_1 \geq \tau_2$ part of the $({\a\b})=(++)$ and $(+-)$ grids, which reduces about half of the computation task. 

\item After obtaining the tabulated grids of $(\tau_1, \tau_2, \mathcal{I}_{\a \b}^{(1)})$ with $({\a\b})=(++)$ and $(+-)$  from Stage (i), we dress $\mathcal{I}_{\a\b}^{(1)}$ with the bulk-to-boundary propagators and other pre-factors as shown in (\ref{eq:vectormaster}). We then interpolate the dressed grids and  integral over $\tau_1$ and $\tau_2$ to get the 1-loop result, $\mathcal{B}_{\a \b}(k_1, k_2, k_3)$, for a particular pair of SK indices $(\a \b)$ and external momentum configuration $(k_1, k_2, k_3)$. Since the time grids are often tightly spaced, we can approximate the time integral by a Riemann sum of the dressed grids to speed up the evaluation. The final result is given by summing over all combinations of SK indices as well as the momentum permutations, i.e., $\mathcal{B} (k_1, k_2, k_3)= \sum_{\a,\b=\pm} \mathcal{B}_{\a \b} (k_1, k_2, k_3) +\text{2 perms} = 2\text{Re} [\mathcal{B}_{++} (k_1, k_2, k_3) + \mathcal{B}_{+-} (k_1, k_2, k_3)] + \text{2 perms}$, where we applied the reality condition in the second equality.  This stage does not require many computational resources and can be finished in a much shorter period compared to Stage (i).
\end{enumerate}

For the scalar or gauge boson loop we considered, the loop integrals in Stage (i) contain products of Whittaker functions (e.g.~Eqs. (\ref{eq:scalarprop}) and~(\ref{eq:vectorprop}))  with complex orders, i.e., $\text{W}_{\kappa, \lambda} (z)$ with $\kappa, \lambda \in \mathbb{C}$.  It is convenient to implement the numerical code in \texttt{Mathematica 12}, which has native support for those functions. An alternative choice is to use \texttt{mpmath}~\cite{mpmath}, a Python library. In App.~\ref{sec:crosscheckA}, we crosscheck results from \texttt{Mathematica 12} code with those from \texttt{mpmath 1.1} code for the CC signal of the tree-level process for complex scalar in an electric field. We find a good agreement between the two. Further relevant details for the numerical evaluation are listed below:
\begin{itemize}
\item{\bf The choice of chemical potential and mass parameter.} The Whittaker functions, $\text{W}_{\mp\ii\tilde \mu, \pm\ii \tilde \nu}(\pm 2\ii k\tau)$, are highly oscillatory when $\wt\mu$ or $\wt\nu$ is large. As a result, numerical evaluations for the loop integral  $\mathcal{I}_{\a \b}^{(1)}$,  in general, takes a longer time as $\tilde \mu$ or $\tilde \nu$ increases. Here we choose $\tilde \mu$ and $\tilde \nu$ to be $\mathcal {O}(1)$. For the Pauli-Villars regulator~(\ref{eq:pauli}), we set the mass parameter $M$ such that $\tilde \nu_\text{reg}$ of the Whittaker functions $\text{W}_{\mp\ii\tilde \mu, \pm\ii \tilde \nu_\text{reg}}(\pm 2\ii k\tau)$ inside $D_M (k, \tau_1, \tau_2)$, satisfies $\tilde \nu_\text{reg} \geq 2 \tilde \nu$.

\item{\bf Range of the magnitude of the external momentum.} We will present the momentum dependence of our results in two ways. One is the squeezed limit ($k\equiv k_1=k_2\gg k_3$), where the CC signal is expected to appear. The other is a more general range of momenta, { the near-equilateral limit} ($k_1 +k_2 \geq k_3$, $k_1 \leq k_3$, $k_2\leq k_3$), which is often adopted in literature for reporting the bispectrum. { Signals with $k_1=k_2 \leq k_3$ in the near-equilateral limit can be viewed as a natural extension of those with $k_1=k_2\gg k_3$ in the squeezed limit.}  By the scale invariance, we can fix $k_3 = 1$ without loss of generality. To set the range of the time grid for Stage (i), we first need to  specify the range of $k_{1,2,3}$ of interest. 
For the squeezed limit, we consider $k \in [1, k_{\max}]$ with $k_{\max} \sim 10^3$, where we expect to see $\mathcal O (\tilde \nu)$ oscillations per decade in $k$. For the near-equilateral limit, we consider the triangle region with $k_{1,2} \in [k_{\min}, 1]$ with $k_{\min} \sim 0.1$.

\item{\bf Range and spacing of the time grid.} In (\ref{eq:vectormaster}), we take $\tau_{1,2} \in (-\infty, 0)$ for the time integral. But for the numerical evaluation, the integral range should be finite. Therefore we need to introduce a finite interval $\tau_{1,2} \in [\tau_i, \tau_f]$. To capture all relevant physics, the initial time $\tau_i$ needs to be set early enough so that all relevant modes are deep inside the horizon ($|k\tau_i| \gg 1$) and well before the particle production ($|k\tau_i| \gg \tilde \mu$).  The final time $\tau_f$ needs to be late enough such that all modes are well outside the horizon ($|k\tau_f| \ll 1$). Given $k \in [k_{\min}, k_{\max}]$ with $k_{\min}\sim 0.1$  and $k_{\max} \sim 10^3$, we choose $\tau_i = -200$ and $\tau_f = - 10^{-5}$ and log-evenly spacing the interval $\tau_{1,2} \in [\tau_i, \tau_f]$ into $N_\text{grid}=440$ pieces. In the end, we get a 2-dim time grid of $(\tau_1, \tau_2)$ with $(N_\text{grid}+1)^2$ grid points in total for each SK indices combination. Given the setup, each grid point occupies the same area in the log space, $[(\ln |\tau_i| -\ln |\tau_f|)/N_\text{grid}]^2$, which we will use as the measure for the Riemann sum in Stage (ii).

\begin{figure}[t]
   \centering
   \includegraphics[width=0.405\textwidth]{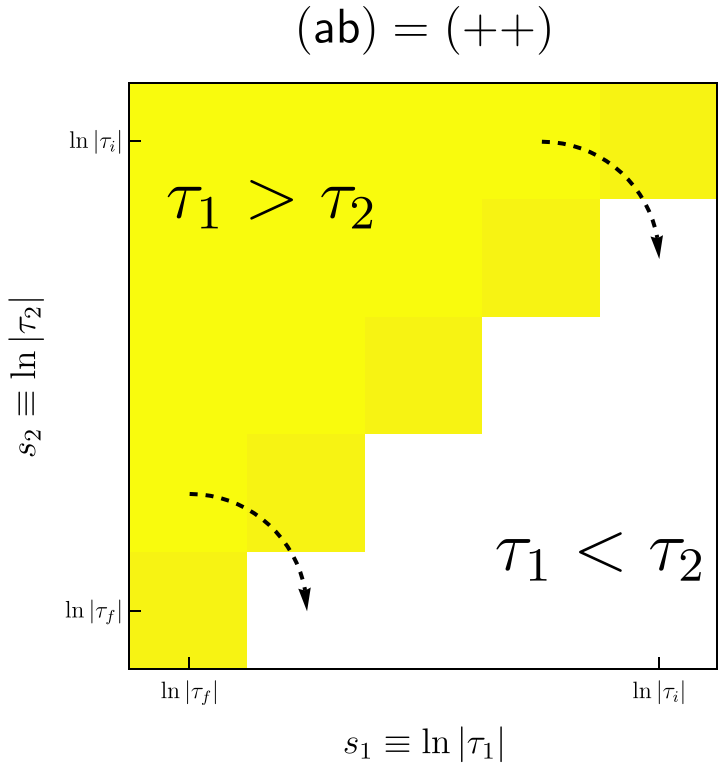}\quad \includegraphics[width=0.4\textwidth]{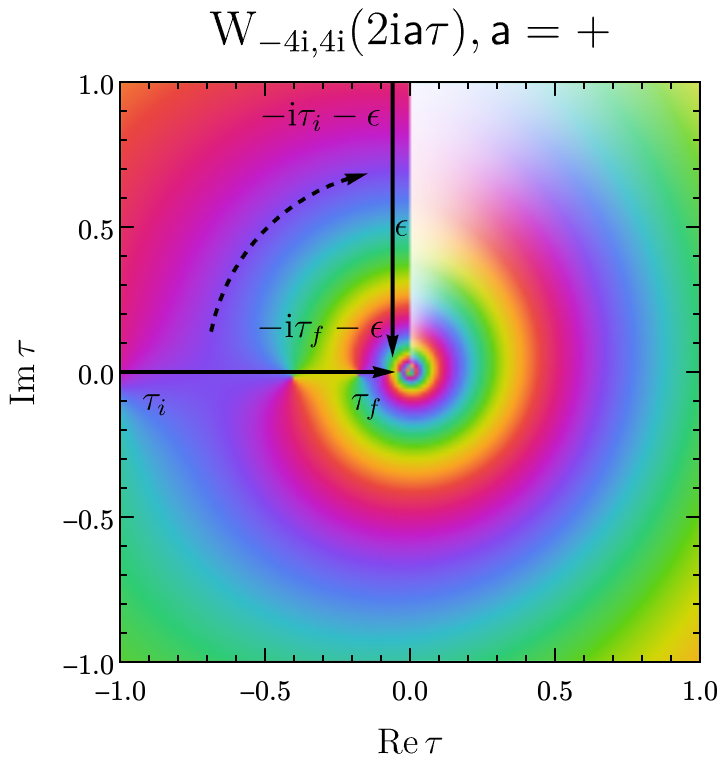}    \caption{(\emph{Left}) 2-dim time grid in $(\ln|\tau_1|, \ln|\tau_2|)$ for the pinched 1-loop diagram with SK indices $(\a \b) =(++)$. We only need to evaluate the loop integrals for the grid with $\tau_1 \geq \tau_2$ (colored part). The loop integral values for the grid with $\tau_1 < \tau_2$ (white part) can be copied from the evaluated grid up to the exchange of $\tau_1 \leftrightarrow \tau_2$. A similar reduction of computational task can be achieved for the $(\a \b) =(+-)$ diagram, where we only need to evaluate the grid with $\tau_1 \geq \tau_2$. See text for more details. (\emph{Right}) An overlay of the integration contour of $\tau$ on top of the complex plot for the Whittaker function $\text{W}_{-4\ii, 4\ii} (2\ii \a \tau)$ with SK index $\a = +$. The color codes the features of the complex function~\cite{complexplot} and the white wedge along the positive imaginary axis represents the branch cut of the Whittaker function. To achieve a better convergence for the time integral while avoiding the branch cut, we rotate $\tau \in [\tau_i, \tau_f]$ to $-\ii \tau \in [-\ii \tau_i -\epsilon, -\ii \tau_f -\epsilon]$, where $\epsilon$ is a small positive real number.}
   \label{fig:whittaker}
\end{figure}

\item {\bf Reducing the grid evaluation for the diagram with $(\a\b) = (++)$.} For both scalar and gauge boson loop diagrams, the key part of the CC signal is given in the form of
\begin{equation}
\mathcal{B}_{\a\b} =  \int_{-\infty}^0 \di \tau_1 \int_{-\infty}^0 \di \tau_2 \int\FR{\di^3\mb q}{(2\pi)^3} \cdots D_{\a \b}( q; \tau_1, \tau_2) D_{\b\a}( p; \tau_2, \tau_1),
\label{eq:simplepp}
\end{equation}
where we only keep track terms of interest. For the $(\a \b) = (++)$ diagram, the loop propagator is a piece-wised function, $D_{++}( q; \tau_1, \tau_2) = D_> ( q; \tau_1, \tau_2) \theta(\tau_1-\tau_2) + D_< ( q; \tau_2, \tau_1) \theta(\tau_2-\tau_1)$, as shown in~\eqref{SKprop}. Using the relation $D_<( q; \tau_1, \tau_2) = D_>( q; \tau_2, \tau_1)$, the integral (\ref{eq:simplepp}) can be expressed as
\begin{align}
\mathcal{B}_{++} ={}&  \int_{-\infty}^0 \di \tau_1 \int_{-\infty}^{\tau_1} \di \tau_2 \int\FR{\di^3\mb q}{(2\pi)^3} \cdots D_{>}( q; \tau_1, \tau_2) D_{>}( p; \tau_1, \tau_2)\nonumber \\
{}&+  \int_{-\infty}^0 \di \tau_1 \int^{0}_{\tau_1} \di \tau_2 \int\FR{\di^3\mb q}{(2\pi)^3} \cdots D_{>}( q; \tau_2, \tau_1) D_{>}( p; \tau_2, \tau_1) \\
={}&  \int_{-\infty}^0 \di \tau_1 \int_{-\infty}^0 \di \tau_2 \int\FR{\di^3\mb q}{(2\pi)^3} \cdots \nonumber \\
& \left[D_{>}( q; \tau_1, \tau_2) D_{>}( p; \tau_1, \tau_2) \theta(\tau_1 -\tau_2) + D_{>}( q; \tau_2, \tau_1) D_{>}( p; \tau_2, \tau_1)  \theta(\tau_2 -\tau_1)\right],
\label{eq:refinepp}
\end{align}
where the two terms inside the square bracket of the last line of~(\ref{eq:refinepp}) are identical with the exchange of $\tau_1 \leftrightarrow \tau_2$. The same symmetry can be found when replacing the propagator $D_{++}$ with the regulated propagator $D_{\text{reg},++}$. Given the symmetry between the two terms, we only need to evaluate $\mathcal{I}_{++}^{(1)}(k_3; \tau_1, \tau_2)$ with $\tau_1 > \tau_2$. After evaluating the loop integrals and obtain the  grid $(\tau_1, \tau_2, \mathcal{I}_{++}^{(1)})$ with $\tau_1 > \tau_2$, one can simply switch $\tau_1$ and $\tau_2$ and copy the loop integral value $\mathcal{I}_{++}^{(1)}$ to tabulate the rest of the grid.\footnote{Note that although $\mathcal{I}_{+-}^{(1)}(k_3, \tau_1, \tau_2)$ shares the same integrand as $\mathcal{I}_{++}^{(1)}(k_3, \tau_1, \tau_2)$ with $\tau_1 < \tau_2$, we could not simply copy the values between the two given their differences in the Wick rotation treatment: we rotate $\tau_{1,2} \to -\ii \tau_{1,2}$ ($\tau_1 \to + \ii \tau_1$ and $\tau_2 \to -\ii \tau_2 $) when evaluating $\mathcal{I}_{++}^{(1)}$ ($\mathcal{I}_{+-}^{(1)}$). See below for Wick rotation.} We illustrate the procedure in the left panel of~\figref{whittaker}.

One subtlety arises for the parameter space just around the diagonal axes, $\tau_1 = \tau_2$. It is the place where the oscillations in the propagators are significant. If we strictly select grid points that satisfy $\tau_1 > \tau_2$, the parameter space will be left out given the finite spacing of the time grid. In the code, we include grid points with $\tau_1 = \tau_2$, illustrated as darker colored squares in the left panel of~\figref{whittaker}, to represent the values of $\mathcal{I}_{++}^{(1)}$ around them.\footnote{ Incidentally, we note that the $\tau_1\simeq\tau_2$ region contributes most to the ``background'' of the correlator and contributes little to the ``signal.'' Therefore one can imagine to approximate the contribution from this region by taking another pinched limit (namely the EFT limit). While this is an interesting point to check numerically in a future work, we must note immediately that this statement is not true in general. For the 4-point correlator with $s$-channel exchange, for example, the $\tau_1=\tau_2$ region can contribute significantly to the signal as well, depending on the choice of external momenta.}

\item{\bf Change variables for the time integral.}
Because the time grids are log-evenly spaced, it is natural to consider $s_{1,2} \equiv \ln |\tau_{1,2}|$ as the integration variables for the time integrals in Stage (ii). Such a change of variable is convenient for the  Riemann sum approach given the measure for each grid point is a fixed value, $\Delta s_1 \Delta s_2 = [(\ln |\tau_i| -\ln |\tau_f|)/N_\text{grid}]^2$. The change of time variables is also helpful for the interpolate-then-integral approach because it improves the sampling of the integrand for the time integral.

\item{\bf Wick rotation of the time variables.} The time integral in Stage (ii) contains a lot of oscillations from the bulk-to-boundary propagators $\partial_\tau G_\pm (k, \tau) \sim e^{\pm \ii k \tau}$, which is numerically difficult to handle. Hence we perform the Wick rotation on $\tau_{1,2}$, for computations in Stage (i) and (ii), to improve numerical convergence in the early-time limit ($\tau \to \tau_i$).\footnote{There is a potential IR problem of full Wick rotation for processes involving mutual cancellation of IR divergences among different diagrams. Our diagrams are free from such problems.} Thanks to the regularization, our loop integrand vanishes in the large loop-momentum limit, as shown in App.~\ref{sec:large-mometum}. We need to be careful to avoid branch cuts in the Whittaker functions as described below. The sign of the rotation, whether $\tau \to + \ii\tau$ or $\tau \to - \ii\tau$, is determined by the $\ii\ep$-prescription, which is practically equivalent to such that the exponential in the bulk-to-boundary propagators are suppressed when $|\tau| \to \infty$, i.e., $e^{\pm \ii k \tau} \to e^{-k |\tau|} = e^{ + k \tau}$. Therefore we take $\tau \to -\ii \tau$ ($\tau \to + \ii \tau$) if $\tau$ is companied with SK index $\a = +$ ($\a = -$). In particular, we rotate $\tau_{1,2} \to -\ii \tau_{1,2}$ ($\tau_1 \to + \ii \tau_1$ and $\tau_2 \to -\ii \tau_2 $) when evaluating the loop integral $\mathcal{I}_{++}^{(1)}$ ($\mathcal{I}_{+-}^{(1)}$).

\item{\bf Avoiding the branch cut.} The Whittaker W function has a branch cut emanating from 0 to $\infty$. In the implementation of \texttt{Mathematica 12} and \texttt{mpmath 1.1}, the branch cut of the Whittaker function $W_{\kappa, \lambda} (z)$ is along the negative real axis of $z$, which should be avoided when performing Wick rotation. This yields a problem for the Whittaker functions with the argument $z = 2\ii \a k \tau$ when the SK index $\a = +$. After performing the Wick rotation described earlier, $z=2\ii \a k \tau = 2\ii k \tau \to z = 2 k \tau \in \mathbb R^-$, which hits the branch cut of the Whittaker function. To avoid the branch cut, we choose to downshift the wick-rotated $z$ by a small imaginary number $2 \ii k \epsilon$ with $\epsilon\in \mathbb R^+$ such that $z=2\ii  k \tau \to z = 2 k \tau - 2 \ii k \epsilon$ (or equivalently $\tau \to -i \tau - \epsilon$). The right panel of \figref{whittaker} is an illustration of the rotation procedure for $\text{W}_{-4\ii, 4\ii} (2\ii \a \tau)$ with $\a=+$. Given the range of momentum and time we considered, we fix $\epsilon=10^{-9}$ and perform the extra shift after rotating $\tau_{1,2}$, i.e., $\tau_{1,2} \to -\ii \tau_{1,2}-\epsilon$, when evaluating $\mathcal{I}_{++}^{(1)}$.\footnote{For $\mathcal{I}_{+-}^{(1)}$, the Wick-rotated $\tau_{1,2}$ do not hit the branch cut under the rotation rule. Hence the extra small shift is not necessary.}

\item {\bf Reducing the grid evaluation for the diagram with $(\a\b) = (+-)$.}  Under complex conjugation, the Whittaker function has the following property
\beq
\left[\text W_{-\ii \tilde \mu, \ii \tilde \nu} (z)\right]^*= \text W_{+\ii \tilde \mu, \ii \tilde \nu} (z)\quad\text{for~}\tilde \mu, \tilde \nu \in \mathbb{R}~\text{and}~z \in \mathbb{R} ^+.
\label{eq:conj}
\eeq
We utilize it to reduce the computational load for the $(\a\b) = (+-)$ diagram. The CC signal for the diagram is given by
\begin{align}
\mathcal{B}_{+ -} ={}&  \int_{-\infty}^0 \di \tau_1 \int_{-\infty}^0 \di \tau_2 \int\FR{\di^3\mb q}{(2\pi)^3} \cdots D_{+-}( q; -\ii\tau_1, \ii \tau_2) D_{-+}( p; \ii \tau_2, - \ii \tau_1)\nonumber\\
={}& \int_{-\infty}^0 \di \tau_1 \int_{-\infty}^0 \di \tau_2 \int\FR{\di^3\mb q}{(2\pi)^3} \cdots D_>( q; \ii \tau_2,  -\ii\tau_1) D_>( p; \ii \tau_2, - \ii \tau_1),
\end{align}
where we rotated the time variables according to the Wick rotation rules described above. For each propagator, we found 
\begin{align}
D_> (k, \ii\tau_2, -\ii\tau_1; \tilde \mu, \tilde \nu) ={}& \cdots \text{W}_{+\ii \tilde \mu, \ii \tilde \nu} (-2 k \tau_2)\text{W}_{-\ii \tilde \mu, \ii \tilde \nu} ( - 2 k \tau_1) \nonumber \\
={}&\left[\cdots \text{W}_{-\ii \tilde \mu, \ii \tilde \nu} (-2 k \tau_2)\text{W}_{+\ii \tilde \mu, \ii \tilde \nu} ( - 2 k \tau_1)\right]^* \nonumber\\
={}&D_> (k, \ii\tau_1, -\ii\tau_2; \tilde \mu, \tilde \nu)^*,
\label{eq:pmid}
\end{align}
where $\cdots$ represents pre-factors that are symmetric under $\tau_1\leftrightarrow \tau_2$. We apply~\eqref{conj} in the second line of~\eqref{pmid}. The same conjugation property is valid for $D_{\text{reg}, >}$. Using~\eqref{pmid}, the signal can be split into two ``symmetric" parts
\begin{align}
\mathcal{B}_{+ -} ={}& \int_{-\infty}^0 \di \tau_1 \int_{-\infty}^0 \di \tau_2 \int\FR{\di^3\mb q}{(2\pi)^3} \cdots \nonumber [D_>( q; \ii \tau_2,  -\ii\tau_1) D_>( p; \ii \tau_2, - \ii \tau_1) \theta(\tau_1 -\tau_2) \\
& + D_>( q; \ii \tau_2,  -\ii\tau_1) D_>( p; \ii \tau_2, - \ii \tau_1) \theta(\tau_2 -\tau_1)] \nonumber \\
={}&\int_{-\infty}^0 \di \tau_1 \int_{-\infty}^0 \di \tau_2 \int\FR{\di^3\mb q}{(2\pi)^3} \cdots \left\{D_>( q; \ii \tau_2,  -\ii\tau_1) D_>( p; \ii \tau_2, - \ii \tau_1) \theta(\tau_1 -\tau_2) \right.\nonumber \\
{}&\left.+ \left[D_>( q; \ii \tau_1,  -\ii\tau_2) D_>( p; \ii \tau_1, - \ii \tau_2)\right]^* \theta(\tau_2 -\tau_1)\right\},
\label{eq:refinepm}
\end{align}
where the two terms inside the curly bracket of the integral~(\ref{eq:refinepm}) are identical up to the exchange of $\tau_1 \leftrightarrow \tau_2$ and the complex conjugation. The same symmetry can be found when replacing $D_>$ with $D_{\text{reg},>}$. As a result, similar to the case for the $(\a\b) = (++)$ diagram, we only need to evaluate $\tau_1 \geq \tau_2$ part of the time grid for the loop integral $\mathcal I_{+-}^{(1)}$. After evaluating $\mathcal I_{+-}^{(1)}$ for $\tau_1 \geq \tau_2$, one can simply switch the position of $\tau_1$ and $\tau_2$ and conjugate the corresponding $\mathcal I_{+-}^{(1)}$  values to tabulate the rest of the time grid.

The first term in the loop integral (\ref{eq:refinepp}) and that of (\ref{eq:refinepm}) can be combined together as a unified function $\text{I}^{(1)} (k_3; t_1, t_2) \equiv (2\pi)^{-3}\int \di^3\mb q \cdots D_> ( q; t_1, t_2)D_> ( p; t_1, t_2)$ with $t_{1,2} = \ii |\tau_{1,2}| -\epsilon$ for $(\a\b)=(++)$ and $t_1 = -\ii |\tau_2|$ and $t_2 = \ii |\tau_1|$ for $(\a\b)=(+-)$. For each grid, we need to evaluate $(N_\text{grid}+1)(N_\text{grid}+2)/2$ grid points. Together $(N_\text{grid}+1)(N_\text{grid}+2)$ grid points are needed to be evaluated in order to get the CC signal for a pinched loop diagram. This is about half of the original total number of grid points, $2(N_\text{grid}+1)^2$.

\item{\bf Integral of the loop momentum.} For the pinched loop diagram, the momentum running in the loop propagators are $\mb q$ and $\mb p \equiv \mb q -\mb k_3$ respectively. We have the freedom to build the coordinate for $\mb q$ and we choose a spherical coordinate with the polar direction ($\theta_{\mb q} = 0$) aligning with the direction of $\mb k_3$. Under such a coordinate, the magnitude of  $\mb p$ is given by $p = \sqrt{q^2 -2   q k_3 \cos \theta_{\mb q} +k_3^2}$. And the integration over $\mb q$ can be expressed in spherical coordinates $\int \di^3 \mb q/(2\pi)^3 \to \int_{-1}^{+1} \di \cos \theta_{\mb q} \int_0^{+\infty} \di q\,  q^2/(2\pi)^2$, where the azimuth angle of $\mb q$ has been integrated out. 

\item{\bf Patched Whittaker function.} It is quite time-consuming to numerically integrate the loop integrand over $q \in [0, \infty)$. To speed up the evaluation, we adopt an approximation for the Whittaker functions, dubbed as the patched Whittaker functions $\ms W_{\kappa, \lambda}(z)$: we replace the Whittaker functions $\text{W}_{\kappa,\lambda} (z)$ with its asymptotic expansion at $z = 0$ and $z=\infty$ for $|z|\ll |\lambda|$ and $|z| \gg |\lambda|$ respectively, while keeping the full function for the intermediate range. To be concrete, we use the third (second) order expansion in $z$ for the region $|z|\ll |\lambda|$ ($|z| \gg |\lambda|$) and patched it at $|z| = |\lambda|/5$ ($|z| = 5|\lambda|$) with the full Whittaker function given our parameter choices, i.e.,

\beq
\label{eq:patchedwhittaker}
\ms{W}_{\kappa, \lambda}(z)  =   \left\{ \begin{matrix} 
     \text{W}^{\text{low}}_{\kappa, \lambda} (z) & |z| <  |\lambda|/5\\
      \text{W}_{\kappa, \lambda} (z) &  |\lambda|/5 \leq |z| < 5 |\lambda|\\
     \text{W}^\text{high}_{\kappa, \lambda} (z) & |z| \geq 5 |\lambda|, \\
   \end{matrix} \right.
\eeq
where
\begin{align}
 \text{W}^{\text{low}}_{\kappa, \lambda} (z) = \frac{z^{\frac{1}{2}-\lambda }}{24}  &\left\{\frac{24 \Gamma (2 \lambda )}{\Gamma \left(-\kappa +\lambda +\frac{1}{2}\right)}\right.\nonumber\\
&+\frac{z \Gamma (2 \lambda -2) }{\Gamma \left(-\kappa +\lambda +\frac{1}{2}\right)} \left[48 \kappa  (\lambda -1)+3 z \left(4 \kappa ^2-2 \lambda +1\right)+\frac{\kappa  z^2 \left(4 \kappa ^2-6 \lambda +5\right)}{2 \lambda -3}\right] \nonumber\\
&+\frac{z^{2 \lambda }  \Gamma (-2 \lambda -3) }{\Gamma \left(-\kappa -\lambda +\frac{1}{2}\right)}\bigg[-48 \left(4\lambda^3+12 \lambda^2 +11 \lambda +3\right)+48 \kappa z \left(2\lambda^2+5\lambda +3\right)\nonumber\\
&\left. \left.
\phantom{\frac{\Gamma}{\Gamma}}-3 z^2(2 \lambda +3)  \left(4 \kappa ^2+2 \lambda +1\right)+\kappa  z^3 \left(4 \kappa ^2+6 \lambda +5\right) \right.\bigg]\right\}
\end{align}
and
\begin{equation}
 \text{W}^\text{high}_{\kappa, \lambda}  (z)= e^{-\frac{z}{2}} z^{\kappa } \left[\frac{-8 \lambda ^2 \left(4 \kappa ^2-8 \kappa +5\right)+16 \lambda ^4+\left(4 \kappa ^2-8 \kappa +3\right)^2}{32 z^2}+\frac{\lambda ^2-\kappa ^2+\kappa -\frac{1}{4}}{z}+1\right].
\end{equation}
In App.~\ref{sec:crosscheckA}, we crosscheck the CC signals from the tree-level diagram for complex scalar in an electric field using the patched Whittaker functions~(\ref{eq:patchedwhittaker}) with those using the full Whittaker functions. We find a good agreement between the full numerical Whittaker and the triply patched Whittaker. Besides we also find a good agreement between the CC signals computed in both ways. The numerical results also agree well with the results from the analytical expression for the $(\a \b) =(+-)$ diagram. 

\item{\bf Contributions from the momentum permutations.} In the above computation, we focus on the case where the loop integral $\mathcal I^{(1)}_{\a\b}$ is controlled by a fixed external momentum $k_3 =1$. To get the contribution from the momentum configuration $(k_1 \leftrightarrow k_3)$ or $(k_2 \leftrightarrow k_3)$, we do not need to re-evaluate the loop integral controlled by $k_1$ or $k_2$. Instead we can utilize the scale invariance of  $\mathcal{S}_{\a\b} (k_1, k_2, k_3)= k_1^2 k_2^2 k_3^2 \mathcal{B}_{\a\b} (k_1, k_2, k_3)$. The contributions from the two momentum permutations, $\mathcal{S}_{\a\b} (k_1 \leftrightarrow k_3)$ and $\mathcal{S}_{\a\b} (k_2 \leftrightarrow k_3)$, are respectively given by
\begin{equation}
\mathcal{S}_{\a\b} (k_1 \leftrightarrow k_3)  = k_1^2 k_2^2 k_3^2 \left(\frac{k_3}{k_1}\right)^6  \mathcal{B}_{\a\b}\left(\frac{k_3^2}{k_1}, \frac{k_2 k_3}{k_1}, k_3\right)
\label{eq:switch1}
\end{equation}
and
\begin{equation}
 \mathcal{S}_{\a\b} (k_2 \leftrightarrow k_3) = k_1^2 k_2^2 k_3^2 \left(\frac{k_3}{k_2}\right)^6 \mathcal{B}_{\a\b}\left(\frac{k_1 k_3}{k_2}, \frac{k_3^2}{k_2}, k_3\right),
 \label{eq:switch2}
\end{equation}
{ where we rescale $k_1$, $k_2$, and $k_3$ of $\mathcal{S}_{\a\b} (k_1 \leftrightarrow k_3)$ ($\mathcal{S}_{\a\b} (k_2 \leftrightarrow k_3)$) by a common factor of $k_3/k_1$ ($k_3/k_2$) in the derivation of \eqref{switch1} (\eqref{switch2}).}
Below we will use $\mathcal S_{\a\b}(k_1, k_2, k_3)$ to represent the sum of the signal from the momentum permutations.
\end{itemize}

Utilizing the above procedures, parameter choices, and special treatments, we find the loop integral at each time grid point in Stage (i) takes a few CPU hours to finish on average. Given a total number of 194,922 grid points (after applying the grid reduction) to evaluate, it costs $\mathcal{O} (10^5)$ CPU hours to get the final CC signal for one pinched loop diagram.

\subsection{Results}

\begin{figure}[t!]
   \centering
   \includegraphics[width=\textwidth]{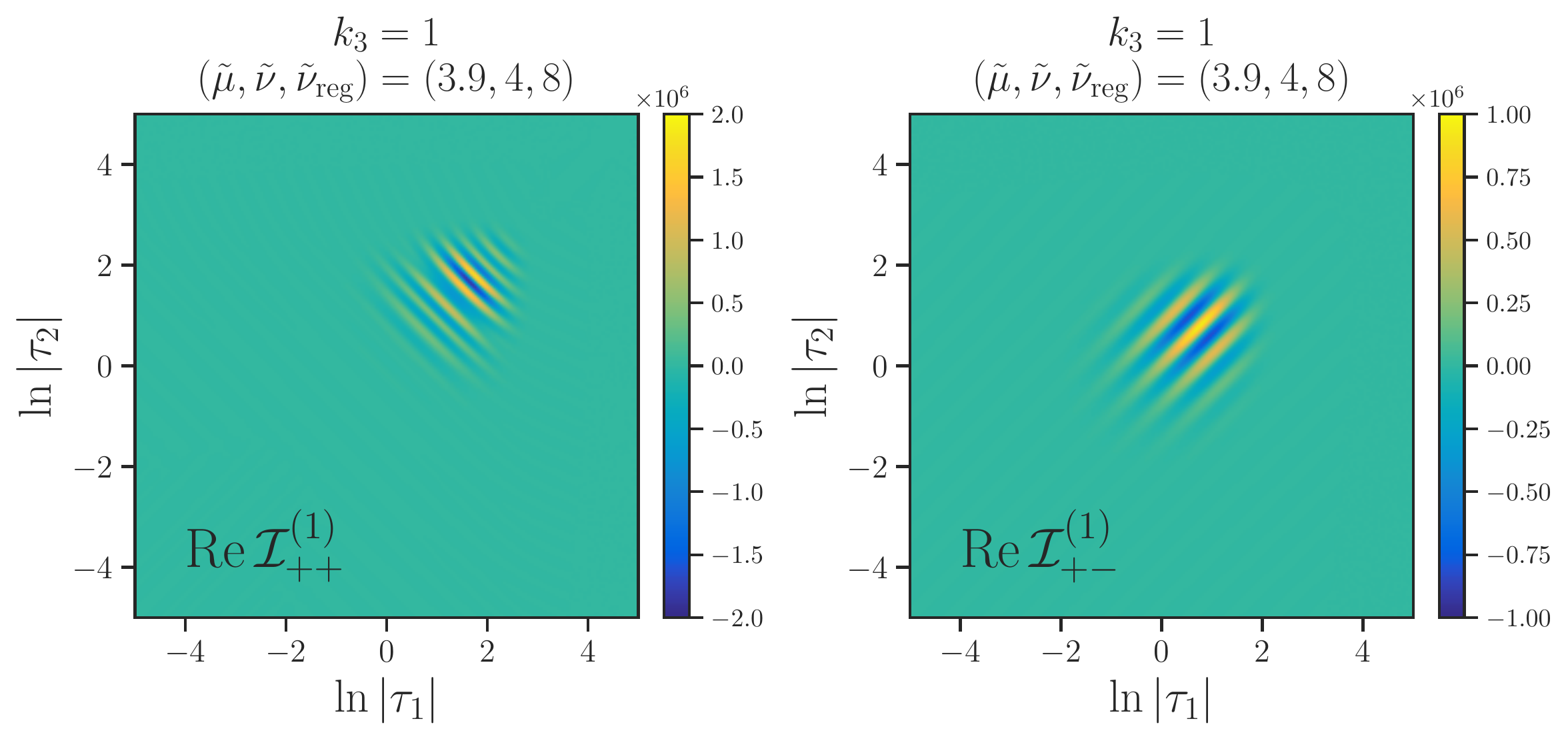}
   \caption{(\emph{Left}) The real part of the loop integral $\mathcal I_{++}^{(1)}$ (part of the full grid) for a pinched 1-loop diagram of scalar with external momentum $k_3=1$ and parameter $(\tilde \mu, \tilde \nu, \tilde \nu_\text{reg}) = (3.9, 4, 8)$. (\emph{Right}) The real part of the loop integral $\mathcal I_{+-}^{(1)}$ (part of the full grid) with the same parameter setting. }
     \label{fig:s_int_1}
\end{figure}

The 1-loop 3-point correlator is the sum of contributions from all SK diagrams, $\mathcal {S} (k_1,k_2,k_3)= \sum_{\a,\b=\pm} \mathcal{S}_{\a \b} = 2 \text{Re}\, \mathcal{S}_{++} + 2 \text{Re}\, \mathcal{S}_{+ -}$, where we applied the reality condition in the last equality. Given the reality of the dressing functions and pre-factors, the value of $\text{Re}\,\mathcal{S}_{++}$ ($\text{Re}\, \mathcal{S}_{+-}$) is determined by the real part of the loop integral $\mathcal I_{++}^{(1)}$ ($\mathcal I_{+-}^{(1)}$). The left and right panels of \figref{s_int_1} respectively shows $\text{Re}\,\mathcal I_{++}^{(1)} (k_3 =1; \ii |\tau_1| -\epsilon, \ii |\tau_2|-\epsilon)$ and $\text{Re}\,\mathcal I_{+-}^{(1)}(k_3 =1; \ii |\tau_1|, -\ii |\tau_2|)$ as a function of $\ln |\tau_1|$ and $\ln |\tau_2|$  (parts of the full grids) for the 1-loop diagram of scalar with parameter $(\tilde \mu, \tilde \nu, \tilde \nu_\text{reg}) = (3.9, 4, 8)$.  The integration results contain oscillatory patterns in both $\text{Re}\,\mathcal I_{++}^{(1)}$ and $\text{Re}\, \mathcal I_{+-}^{(1)}$. For the former case, the oscillations are along the axes of $|\tau_1| = |\tau_2|$.  And for the latter case, the oscillations appear perpendicular to the axes of $|\tau_1| = |\tau_2|$. For both cases, the oscillations become the most significant in the region around $\ln |\tau_1| = \ln |\tau_2| \approx 2$.

\begin{figure}[t!]
   \centering
   \includegraphics[width=\textwidth]{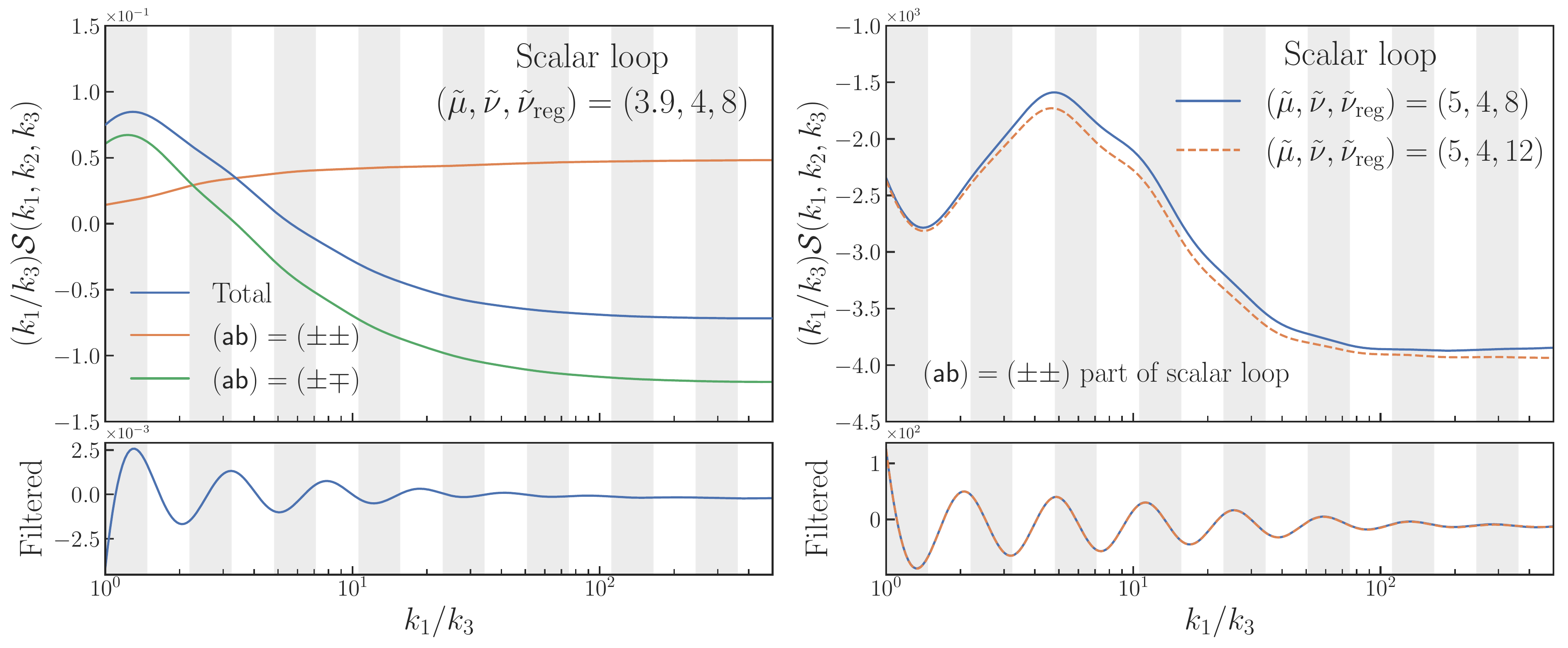}
   \caption{(\emph{Left}) The rescaled CC signal, $(k_1/k_3) \mathcal S(k_1,k_2,k_3)$, in the squeezed limit, $k_1=k_2 \gg k_3$, for the 1-loop diagram of scalar with $(\tilde \mu, \tilde \nu, \tilde \nu_\text{reg}) = (3.9, 4, 8)$ (blue line). Contributions from the diagrams with SK indices of the same (opposite) sign are shown as the orange (green) line. The lower sub-panel shows the filtered CC signal where its non-oscillatory component is attenuated. (\emph{Right}) The CC signal from diagrams with SK indices of the same sign, $ \mathcal S_{++} + \mathcal S_{--} = 2 \text{Re}\,\mathcal S_{++}$, rescaled by $k_1/k_3$, in the squeezed limit, $k_1=k_2 \gg k_3$, for a pinched 1-loop diagram of scalar with parameter $(\tilde \mu, \tilde \nu, \tilde \nu_\text{reg}) = (5, 4, 8)$ (solid blue line) and parameter $(\tilde \mu, \tilde \nu, \tilde \nu_\text{reg}) = (5, 4, 12)$ (dashed orange line). The lower sub-panel shows filtered signals for the two sets of parameters, where their non-oscillatory components are attenuated. A set of gray bands, which are evenly separated by $\Delta \log_{10} (k_1/k_3) = \pi (\log_{10} e)/\tilde \nu$, are included for all the panels as a guide to the oscillatory pattern.}
   \label{fig:s_1}
\end{figure}

By taking the time integral (Riemann sum) over the grids of the dressed loop integral and summing over the momentum permutations, we obtain the CC signal for a given momentum configuration $(k_1,k_2,k_3)$. In the left panel of~\figref{s_1}, we show the resulting signal, rescaled by $k_1/k_3$, in the squeezed limit, $k_1=k_2 \gg k_3$. The resulting signal demonstrates periodic oscillations as a function of the logarithm of the momentum ratio, $\log_{10} (k_1/k_3)$, with a period of $T = \pi \log_{10} e/ \tilde \nu$. We insert a set of gray bands, which are evenly separated by $\Delta \log_{10} k_1/k_3 = T$, in the panel as a guide to the pattern. Such oscillations can be also observed from the results of the diagrams with the same(opposite)-sign SK indices, $2\text{Re}\, \mathcal S_{++}$ ($2\text{Re}\, \mathcal S_{+-}$), that contribute to the final signal. To see the oscillatory pattern more clearly, we attenuate  the smooth component of the signal. In the lower sub-panel, we show the filtered results by  passing the signal through a high-pass filter with a Gaussian window function and a cut-off frequency of $\omega_c = 0.05 \tilde \nu/ (\pi \log_{10} e)$. The amplitude of the wiggle is about one to two orders of magnitude smaller than that of the full signal. We apply the same high-pass filter for all the filtering procedures in the main text and crosscheck it with other filtering methods in App.~\ref{sec:crosscheckB}.

\begin{figure}[t!]
   \centering
 \includegraphics[width=0.96\textwidth]{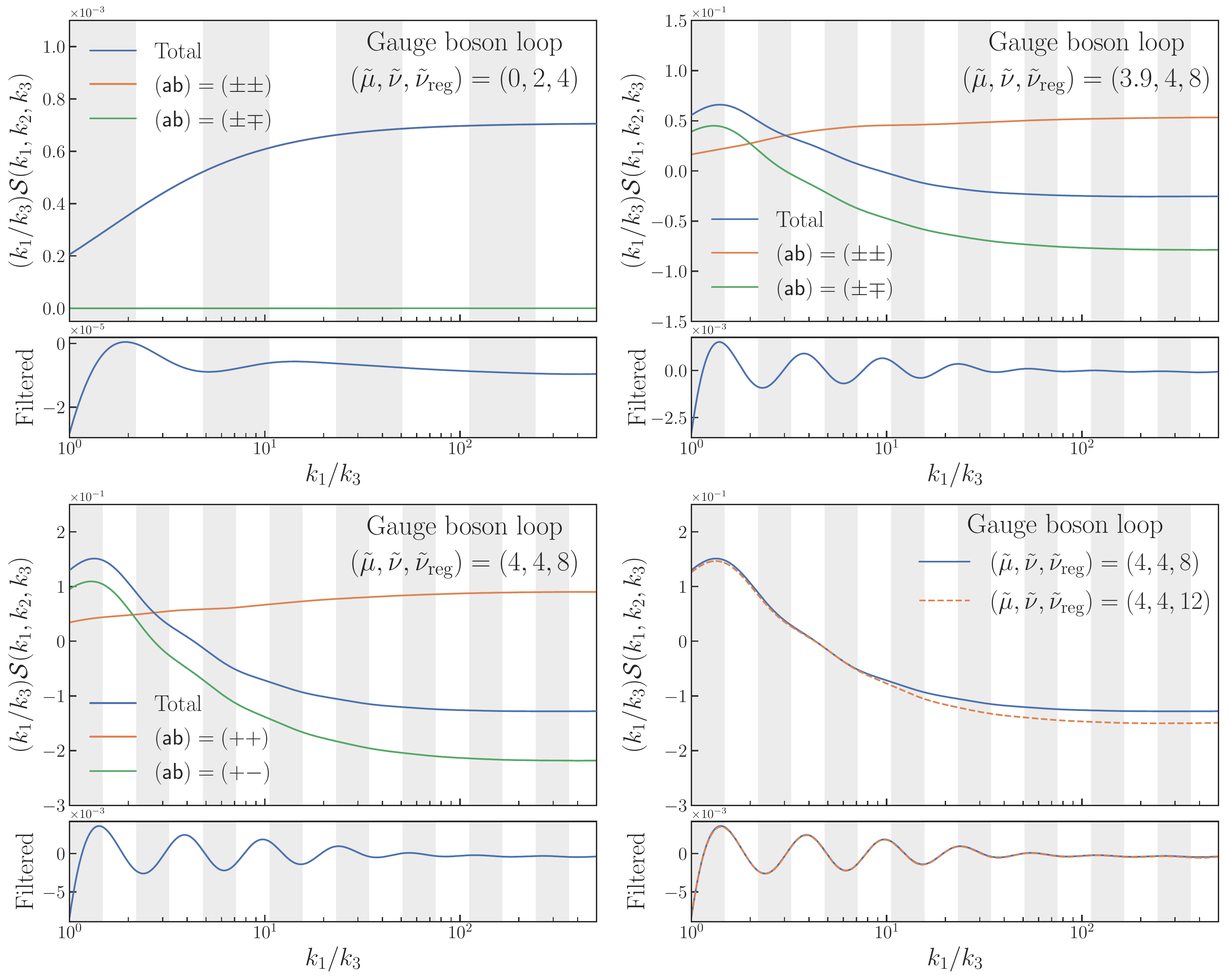}
    \caption{(\emph{Upper left}) The rescaled CC signal, $(k_1/k_3)\mathcal S(k_1,k_2,k_3)$, in the squeezed limit, $k_1=k_2 \gg k_3$, for the 1-loop diagram of gauge boson without chemical potential, where the parameter is set by $(\tilde \mu, \tilde \nu, \tilde \nu_\text{reg}) = (0, 2, 4)$ (blue line). Contributions from the diagrams with SK indices of the same (opposite) sign are shown as the orange (green) line. The lower sub-panel shows the filtered CC signal where its non-oscillatory component is attenuated (same for the other panels).  (\emph{Upper right}) The CC signal for the 1-loop gauge boson with chemical potential, where the parameter is set by $(\tilde \mu, \tilde \nu, \tilde \nu_\text{reg}) = (3.9, 4, 8)$. (\emph{Lower left}) The CC signal for the same diagram with $(\tilde \mu, \tilde \nu, \tilde \nu_\text{reg}) = (4, 4, 8)$. (\emph{Lower right}) The total CC signal for the same diagram with $(\tilde \mu, \tilde \nu, \tilde \nu_\text{reg}) = (4, 4, 12)$ (dashed orange) compared with that of $(\tilde \mu, \tilde \nu, \tilde \nu_\text{reg}) = (4, 4, 8)$ (solid blue). A set of gray bands, which are evenly separated by $\Delta \log_{10} (k_1/k_3) = \pi (\log_{10} e)/\tilde \nu$, are included for all the panels as a guide to the oscillatory pattern.}
   \label{fig:v_2}
\end{figure}

Next, we look into the dependence of the signal on the loop regulator, $\nu_\text{reg}$. The right panel of~\figref{s_1} shows the resulting signal from the diagrams with the same-sign SK indices, $ \mathcal S_{++} + \mathcal S_{--} = 2 \text{Re}\,\mathcal S_{++}$, in the squeezed limit. The solid blue and dashed orange curves are results with parameter $(\tilde \mu, \tilde \nu, \tilde \nu_\text{reg}) = (5, 4, 8)$ and $(\tilde \mu, \tilde \nu, \tilde \nu_\text{reg}) = (5, 4, 12)$ respectively. The resulting signals from different values of $\nu_\text{reg}$ share the same periodicity and phase, and a mild difference in their amplitudes. This indicates the divergence of the loop diagram is logarithmic. The two signals are almost identical after passing through the same high-pass filter.

\begin{figure}[t!]
   \centering
   \includegraphics[width=0.96\textwidth]{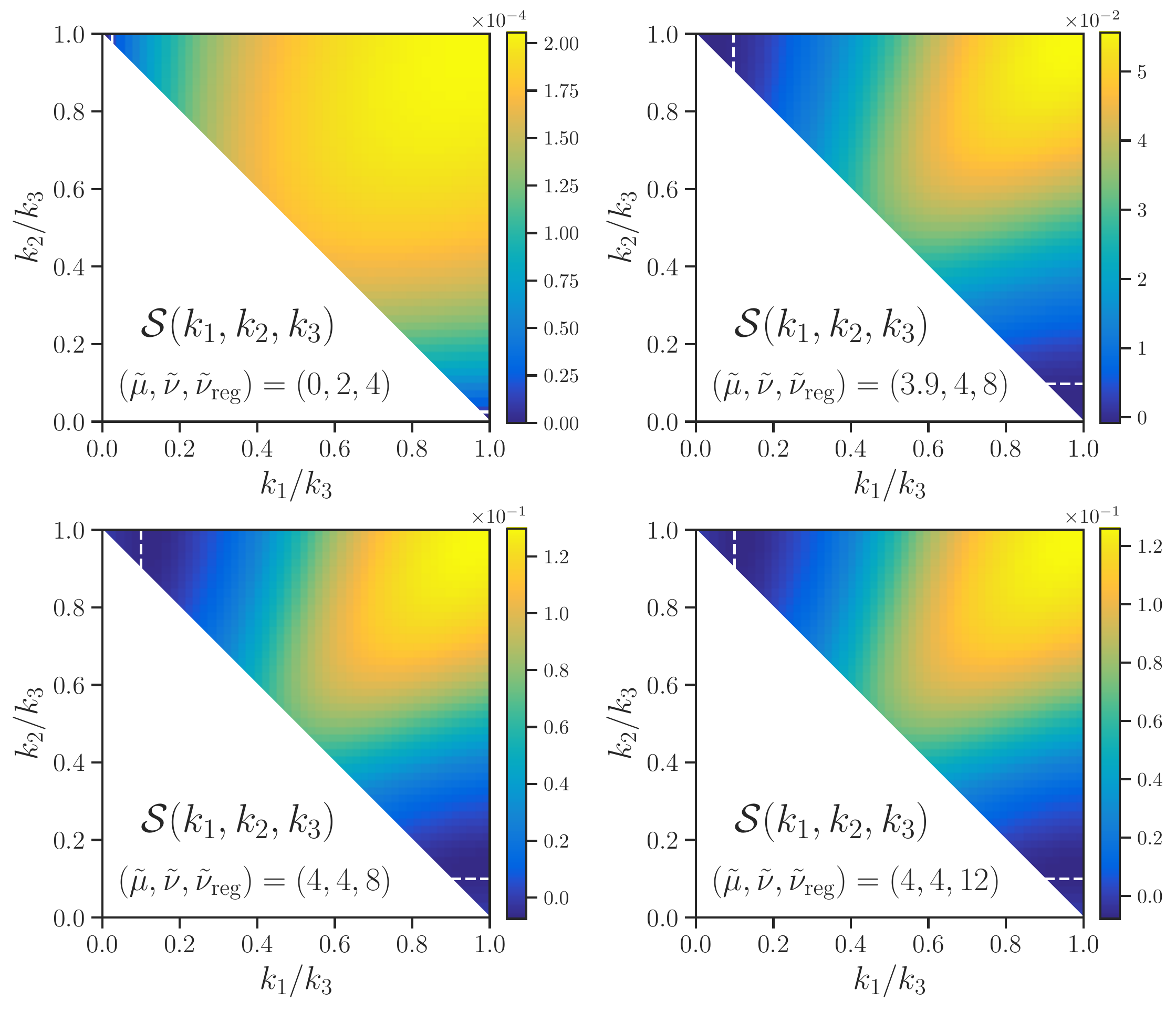}
   \caption{The CC signal $\mathcal S(k_1,k_2,k_3)$ in the near-equilateral limit for the 1-loop diagram of gauge boson with parmeter $(\tilde \mu, \tilde \nu, \tilde \nu_\text{reg}) = (0, 2, 4)$ (\emph{upper left}), $(3.9, 4, 8)$ (\emph{upper right}), $(4, 4, 8)$ (\emph{lower left}), and $(4, 4, 12)$ (\emph{lower right}). The dashed lines mark the boundaries for the region $ k_{1,2}|\tau_i| \gg \max(\tilde \mu, 1)$ when $k_3=1$ and $\tau_i = -200$.}
   \label{fig:neq}
\end{figure}

Now we switch to the gauge boson loop. The upper left panel of~\figref{v_2} shows the signal for the 1-loop diagram of gauge boson without chemical potential, where the parameter is set to $(\tilde \mu, \tilde \nu, \tilde \nu_\text{reg}) = (0, 2, 4)$. The oscillatory signal is relatively tiny since the chemical potential is absent. (Similar trends are observed for the scalar loop case.) Contributions from the diagrams with the opposite-sign SK indices become largely suppressed compared to those from the same-sign SK indices. Once the chemical potential turns on, because the gauge boson signal is proportional to $\exp(-2 \pi h \tilde \mu)$ where $h=+$($-$) for its positive (negative) helicity component, the signal from the negative helicity becomes dominant for $\tilde \mu \gtrsim \mathcal O (1)$. In the upper right panel of~\figref{v_2}, we show the resulting signal in the squeezed limit for the gauge boson with  parameter $(\tilde \mu, \tilde \nu, \tilde \nu_\text{reg}) = (3.9, 4, 8)$. The oscillatory pattern is obvious in the filtered signal.  The shape and amplitude of the (filtered) signal are similar to those of the scalar loop shown in the left panel of~\figref{s_1}. Unlike the case without chemical potential, contributions from the diagrams with the same/opposite-sign SK indices are comparable. The lower left panel of~\figref{v_2} shows the results for a slightly different parameter choice $(\tilde \mu, \tilde \nu, \tilde \nu_\text{reg}) = (4, 4, 8)$, where we found the signal is generally a factor of $\exp(2\pi \Delta \tilde \mu)\sim 1.8$ larger than that with $(\tilde \mu, \tilde \nu, \tilde \nu_\text{reg}) = (3.9, 4, 8)$.  We demonstrate the $ \nu_\text{reg}$-dependence in the lower right panel, where the  solid blue and  dashed orange lines are the resulting total signals with parameter  $(\tilde \mu, \tilde \nu, \tilde \nu_\text{reg}) = (4, 4, 8)$ and $(\tilde \mu, \tilde \nu, \tilde \nu_\text{reg}) = (4, 4, 12)$ respectively. Similar to the scalar scenario, the two signals share a mild difference in their amplitudes and yield almost identical oscillation patterns after passing through the same high-pass filter.

In~\figref{neq}, we plot the 3-point correlator mediated by the gauge boson loop for more general momentum configurations. The gauge boson loop is dominated by the negative helicity mode for large chemical potentials. We show the four sets of the parameter $(\tilde \mu, \tilde \nu, \tilde \nu_\text{reg})$ described earlier. We use the dash lines to mark the boundaries of the regions with $k_{1,2} |\tau_i| \gg \max(\tilde \mu, 1)$,\footnote{To be concrete, the boundaries are set by $k_{1,2} |\tau_i| =5 \max(\tilde \mu, 1)$. } where the numerical results are not handicapped by the finite value of $\tau_i$ during the time integral.

\subsection{Comparison with Analytic Estimates}

 \begin{figure}[t!]
   \centering
   \includegraphics[width=0.8\textwidth]{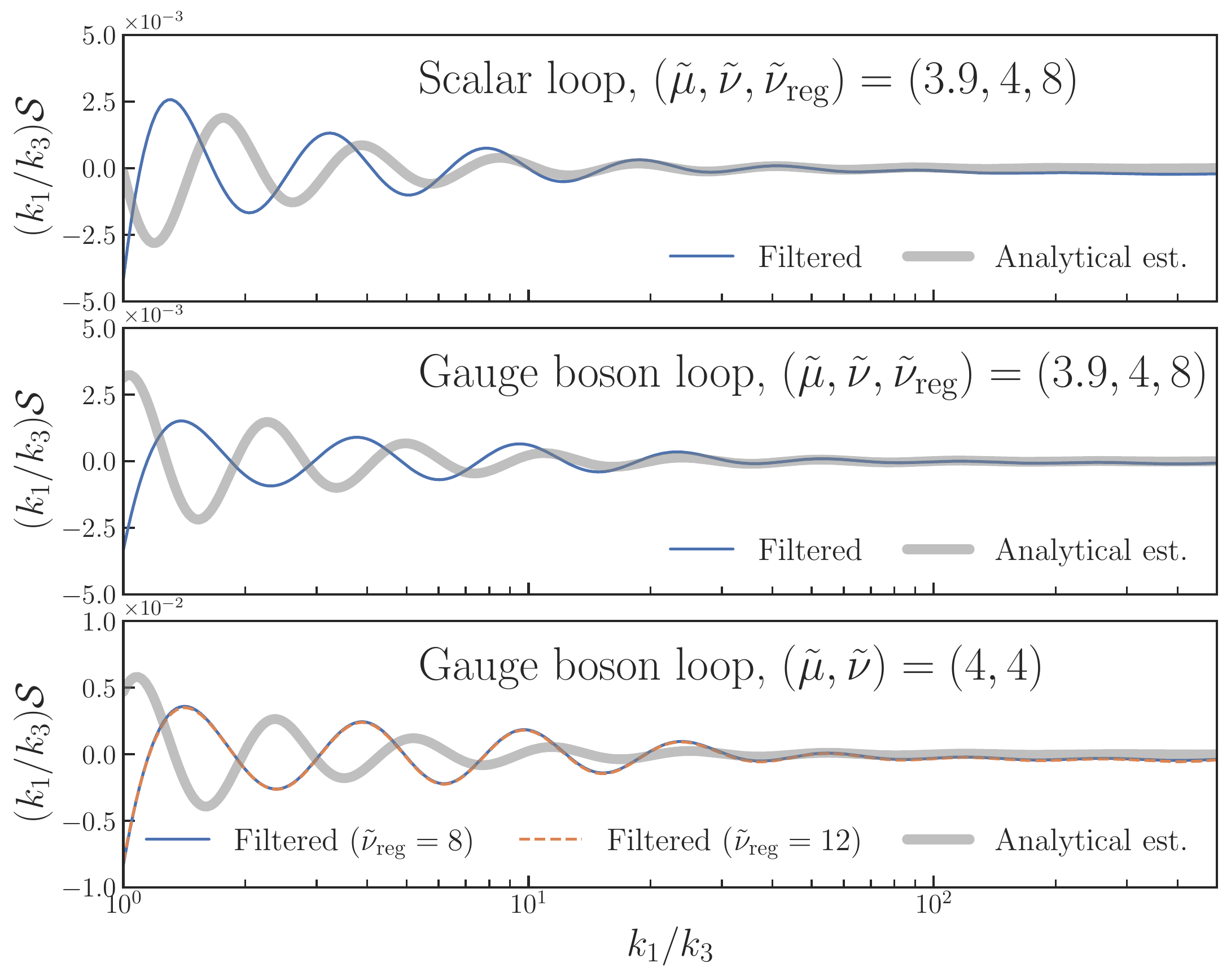}
   \caption{ (\emph{Upper}) A comparison between the filtered CC signal (blue) and its analytical estimation (gray). The filtered CC signal is from the 1-loop diagram of scalar with parameter $(\tilde \mu, \tilde \nu, \tilde \nu_\text{reg}) = (3.9,4,8)$. Both signals are rescaled by $k_1/k_3$. (\emph{Middle}) A similar comparison for the 1-loop diagram of gauge boson with parameter $(\tilde \mu, \tilde \nu, \tilde \nu_\text{reg}) = (3.9,4,8)$. (\emph{Lower}) A similar comparison for  the 1-loop diagram of gauge boson with  parameter $(\tilde \mu, \tilde \nu, \tilde \nu_\text{reg}) = (4,4,8)$ (solid blue) or $(\tilde \mu, \tilde \nu, \tilde \nu_\text{reg}) = (4,4,12)$ (dashed orange).}
   \label{fig:analytical_est}
\end{figure}

The loop amplitudes calculated above have been estimated in previous works. The estimates usually assumed a late-time expansion of the loop propagator which is not entirely valid for 3-point functions. Explicitly, it was estimated that the signal part of a 1-loop process mediated by a boson with mass $\wt m\equiv m/H\gg 1$ and chemical potential $\wt\mu\equiv \mu/H$ takes the following form as a function of $\varrho\equiv k_3/k_1$,
\begin{align}
\label{analest}
  \lim_{\varrho\to 0}\mathcal{S}_\text{signal}^\text{(est.)}(\varrho)\simeq \FR{1}{16\pi^2}\wt\mu^y \wt m^z e^{2\pi(\wt\mu-\wt \nu)}\varrho^2\sin(2\wt\nu\log\varrho+\varphi),
\end{align}
where $\wt\nu=\sqrt{\wt m^2-9/4}$ for scalar and $\wt\nu=\sqrt{\wt m^2-1/4}$ for spin-1 gauge bosons. The phase $\varphi$ can be calculated and depends on  $\wt\mu$ and $\wt\nu$ in complicated ways. The powers $y$ and $z$ depend on interactions and are difficult to estimate precisely. It is known that late-time expansion of the intermediate propagators can capture the exponential dependence $e^{2\pi(\wt\mu-\wt\nu)}$ correctly but fails to get the correct power dependence $\wt\mu^y \wt m^z$. But given that the most sensitive dependence comes from the exponential factor, it is still useful to compare our numerical results with the analytical estimates. In~\figref{analytical_est}, we superimpose the result of analytical estimates (\ref{analest}) with the filtered numerical results for four 1-loop diagrams we considered, setting $y=z=0$, and tuning the phases $\varphi$ to match the peaks and valleys of the analytical and numerical oscillatory signals for the region $k_1/k_3 \gg 10$. We found that the overall amplitudes of the numerical results and analytical estimations are compatible. We also observe a small difference in the oscillation frequencies between the two. Our numerical results show a slightly smaller frequency than analytical estimates $\tilde \omega = 2\tilde \nu/\log_{10} e$ in all examples. We are currently unaware of the origin of this small discrepancy. Partly due to this difference and partly due to the failure of analytical estimates at small momentum ratios, the two results disagree in phases at lower values of $k_1/k_3$, while they are in reasonable agreement for $k_1/k_3 \gtrsim 20$, as shown in \figref{analytical_est}.

\section{Conclusions}
\label{sec_conclusions}
We performed a systematic study of the numerical evaluation of the inflation correlators at the 1-loop level. In particular, we presented the first numerical results for a set of 3-point 1-loop processes with spin-0 or -1 particles running in the loop. Without applying crude approximations, our numerical results offer a set of precise templates for the 1-loop process after imposing appropriate renormalization conditions. Our results include both ``signal" and ``background" parts,  that can be used in searches with observational data. While the ``background" can be several orders of magnitude larger, we show that the signal can be separated out by applying high-pass filters. We have also compared our numerical results with analytic estimates adopted in the earlier literature. The amplitudes obtained by both approaches are compatible. At the same time, there is a shift in the frequency   of the oscillatory part for $k_1/k_3$ at a several-percent level. Further clarifying the origin of this shift could be a fruitful future direction to pursue.

As shown in this paper, the numerical implementation of the relevant 1-loop integrals turns out to be highly nontrivial. Many subtleties could appear along with the calculation, including the appropriate choices of integral range and gridding, the highly oscillatory integrand in certain parameter regions, the divergence of the integral and its regularization, the analytical structure of special functions that is relevant when performing Wick rotation, etc. We have spelled out all these subtleties in the paper, which could be useful for more extensive numerical studies in the future.  We stress that our method does not rely on approximations that are often used in previous analytical studies and work usually only in several limiting momentum configurations. Our method thus applies to arbitrary momentum configurations. 

{ It would be interesting to extend our current project to other interesting situations, such as the fermionic 1-loop process, more general triangular 1-loop processes (without taking pinched limit), and more general couplings (e.g.\ $\phi F\wt F$ mentioned in the paper).} It is also straightforward to apply our method to 1-loop mediated trispectrum (4-point correlation functions). The results of these studies can be a useful check of previous analytical estimates and are also useful for building templates for future observations of the 3-point functions.

As the first attempt of numerical evaluation of 1-loop bispectrum, our method is essentially a direct implementation of Feynman integrals from first principles. As we have shown, the evaluation of relevant Feynman integrals is computationally heavy. It would be inefficient to apply our method directly for parameter scanning and templates building today. For these purposes, it would be desirable to look for both analytical simplifications and better numerical strategies. We leave these directions for future studies.

\paragraph{Acknowledgment.} We thank Qianhang Ding, Junwu Huang, Edward W. Kolb, and Yi Wang  for useful discussions. We thank Xingang Chen, Soubhik Kumar, Wayne Hu, Junwu Huang, Hayden Lee, Gustavo Marques-Tavares, and Yi Wang for useful comments on a draft of this manuscript. This work was completed with resources provided by the University of Chicago’s Research Computing Center. We also thank Haipeng An, Lincoln Bryant, Rob Gardner, Pascal Paschos, and Daneng Yang for providing computational resources and technical supports at the early stage of this work. LTW and YZ acknowledge the Aspen Center for Physics for hospitality during the final phase of this study, which was supported by National Science Foundation grant PHY-1607611 and partially supported by a grant from the Simons Foundation. LTW is supported by the DOE grant DE-SC0009924. ZZX is supported by Tsinghua University Initiative Scientific Research Program.  YZ is supported by the Kavli Institute for Cosmological Physics at the University of Chicago through an endowment from the Kavli Foundation and its founder Fred Kavli.

\appendix

\section{Crosscheck numerical implementations with the tree-level diagram for complex scalar in an electric field}
\label{sec:crosscheckA}

Ref.~\cite{Chua:2018dqh}  investigated the Schwinger effect of charged complex scalars under an electric field during inflation. The CC signal for the tree-level diagrams are given by
\begin{align}
\mathcal B(k_1, k_2, k_3; \t \mu, \t \nu)=\sum_{\a \b}\a \b (\ii)^2 &\int_{-\infty}^0 \di\tau_1 \int_{-\infty}^0 \di\tau_2\, \f{1}{(-H \tau_1)^2}\f{1}{(-H \tau_2)^3} \nonumber\\
& \partial_{\tau_1} G_{\a} ( k_1, \tau_1)\partial_{\tau_1} G_{\a} ( k_2, \tau_1) \partial_{\tau_2} G_{\b} ( k_3, \tau_2) {D_{\b \a} ( k_3; \tau_2, \tau_1; \t \mu, \t \nu)},
\end{align}
which contains time integral over $\tau_{1,2}$ and the Whittaker functions in the bulk-to-bulk propagator $D_{\b \a}(k_3; \tau_2, \tau_1)$ where $\tilde \mu$ and $\tilde \nu$ are the electric field strength parameter and the mass parameter respectively.\footnote{ Ref.~\cite{Chua:2018dqh} uses $\kappa$ and $\mu$ to represent the electric field strength parameter and the mass parameter respectively. They are related to  $\tilde \mu$ and $\tilde \nu$ by $\kappa = - \ii\tilde \mu$ and $\mu = \ii \tilde \nu$.} Numerical evaluations of the diagram take a relatively short time (for codes with the full Whittaker  and the patched Whittaker functions) while some treatments, such as the Wick rotation of $\tau_{1,2}$ and avoiding the branch cut of the Whittaker functions, are similar to those for the loop diagram. In addition, we have an analytical expression for the signal of the $(\a\b) = (+-)$ diagram to compare with the numerical results,
\beq
{\mathcal B_{+-}(k_1, k_2, k_3; \t \mu, \t \nu) = \frac{\pi ^2 e^{ \pi  \t \mu } \left(16 \t \nu ^4 + 40 \t \nu ^2+9\right)}{1024\, \Gamma (1+\ii \t \mu )\, k_1 k_2 k_3^4 }  \sec ^2(\pi  \lambda ) \, _2\tilde{F}_1\left(-\ii \t \nu+\frac{5}{2} ,\ii \t\nu +\frac{5}{2};-\ii \t \mu +3;\frac{k_3 -k_1 -k_2}{2 k_3}\right)},
\label{eq:myany}
\eeq
where $_2\tilde{F}_1$ is the regularized hypergeometric function.
Therefore, the tree-level diagram for complex scalars in an electric field provides an ideal scenario for us to crosscheck the convergence of various numerical implementations described in Sec.~\ref{sec_setup}. We first compare numerical results from those using the full Whittaker functions to those using the patched Whittaker functions. We then compare the numerical results from coding in $\texttt{Mathematica}$ to those from coding in \texttt{mpmath}. 

\subsection{Whittaker function vs. patched Whittaker function}
\begin{figure}[t!]
   \centering
\includegraphics[width=0.96\textwidth]{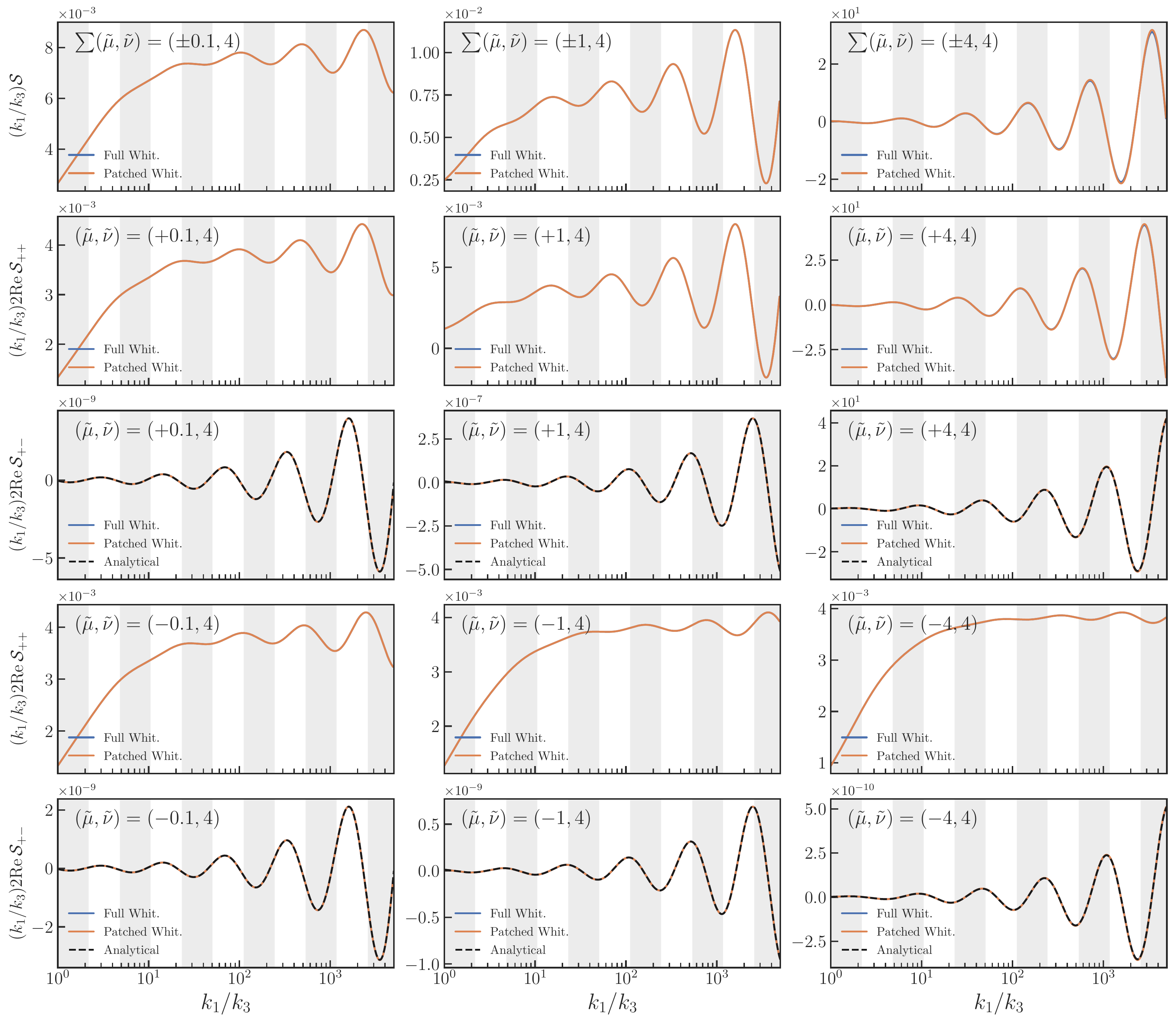} 
\caption{(\emph{Row 1}) The  CC signal $\mathcal S$, rescaled by $k_1/k_3$, in the squeezed limit, $k_1=k_2 \gg k_3$, for a tree-level diagram for complex scalar in an electric field, where $\mathcal S = 2 \text{Re} [\mathcal{S}_{++} (\tilde \mu, \tilde \nu) +\mathcal{S}_{+-} (\tilde \mu, \tilde \nu) + \mathcal{S}_{++} (-\tilde \mu, \tilde \nu) +\mathcal{S}_{+-} (-\tilde \mu, \tilde \nu) ]$. The electric field strength and the mass parameter are set such that $(\tilde \mu, \tilde \nu)=(\pm 0.1,4), (\pm 1,4), (\pm 4,4)$ for \emph{column 1-3} respectively. We compare numerical results from the full Whittaker functions (blue lines) and those from the patched Whittaker functions (orange lines). (\emph{Row 2-5}) The four components that contribute to $\mathcal S$. We also include the analytical results (dashed black) for $\mathcal{S}_{+-}$ components to compare with the numerical results. A set of gray bands, which are evenly separated by $\Delta \log_{10} (k_1/k_3) = 2 \pi (\log_{10} e)/\tilde \nu$, are included for all panels as a guide to the oscillatory pattern. }
\label{fig:fwvspw}
\end{figure}

We consider the CC signal, ${\mathcal S}(k_1, k_2, k_3)$, in the squeezed limit $k_1=k_2 \gg k_3$, with the electric field strength and the mass parameter $(\tilde \mu, \tilde \nu)=(0.1,4)$, $(1,4)$, and $(4,4)$. For a given set of parameter, the total signal is the sum of positive and negative electric field strength parameter, i.e., $\mathcal S = \sum_{\a \b} \mathcal S_{\a\b}(\tilde \mu, \tilde \nu) + \sum_{\a \b} \mathcal S_{\a\b}(-\tilde \mu, \tilde \nu) =  \sum \mathcal S_{\pm \pm}(\tilde \mu, \tilde \nu)   +  \sum \mathcal S_{\pm \mp}(\tilde \mu, \tilde \nu)  + (\tilde \mu \to -\tilde \mu)= 2 \text{Re} [\mathcal{S}_{++} (\tilde \mu, \tilde \nu) +\mathcal{S}_{+-} (\tilde \mu, \tilde \nu) ] + (\tilde \mu \to - \tilde \mu)$, where we apply the reality condition of SK diagrams in the last equality. The final signals, rescaled by $k_1/k_3$, are shown in the first row of~\figref{fwvspw}, where the blue lines are the results from the numerical evaluation using the full Whittaker functions and the orange lines are the results using the patched Whittaker functions (\ref{eq:patchedwhittaker}). We see a good agreement between the two implementations. Such agreement can be also seen at the level of component that contributes to the signal, as shown in rows 2--5 of~\figref{fwvspw}. Note that for $\sum \mathcal S_{\pm \mp} = 2 \text{Re}\,\mathcal S_{+-}$ components (row 3 and 5), we also include results from the analytical expression of (\ref{eq:myany}), which are in good agreement with the numerical results.

All the CC signals show oscillatory patterns with a periodicity of $T = 2\pi \log_{10}e /\tilde \nu\approx 0.68$ with respect to $\log_{10} (k_1/k_3)$. For each set of parameters, the dominant contributions are from the diagrams with SK indices of the same sign. As $\tilde \mu$ increases, the overall size of the signal increases, and the oscillatory patterns of the signals begin to dominate  over the smooth backgrounds.

\subsection{\texttt{Mathematica} vs. \texttt{mpmath}}
\begin{figure}[t!]
   \centering
\includegraphics[width=0.8\textwidth]{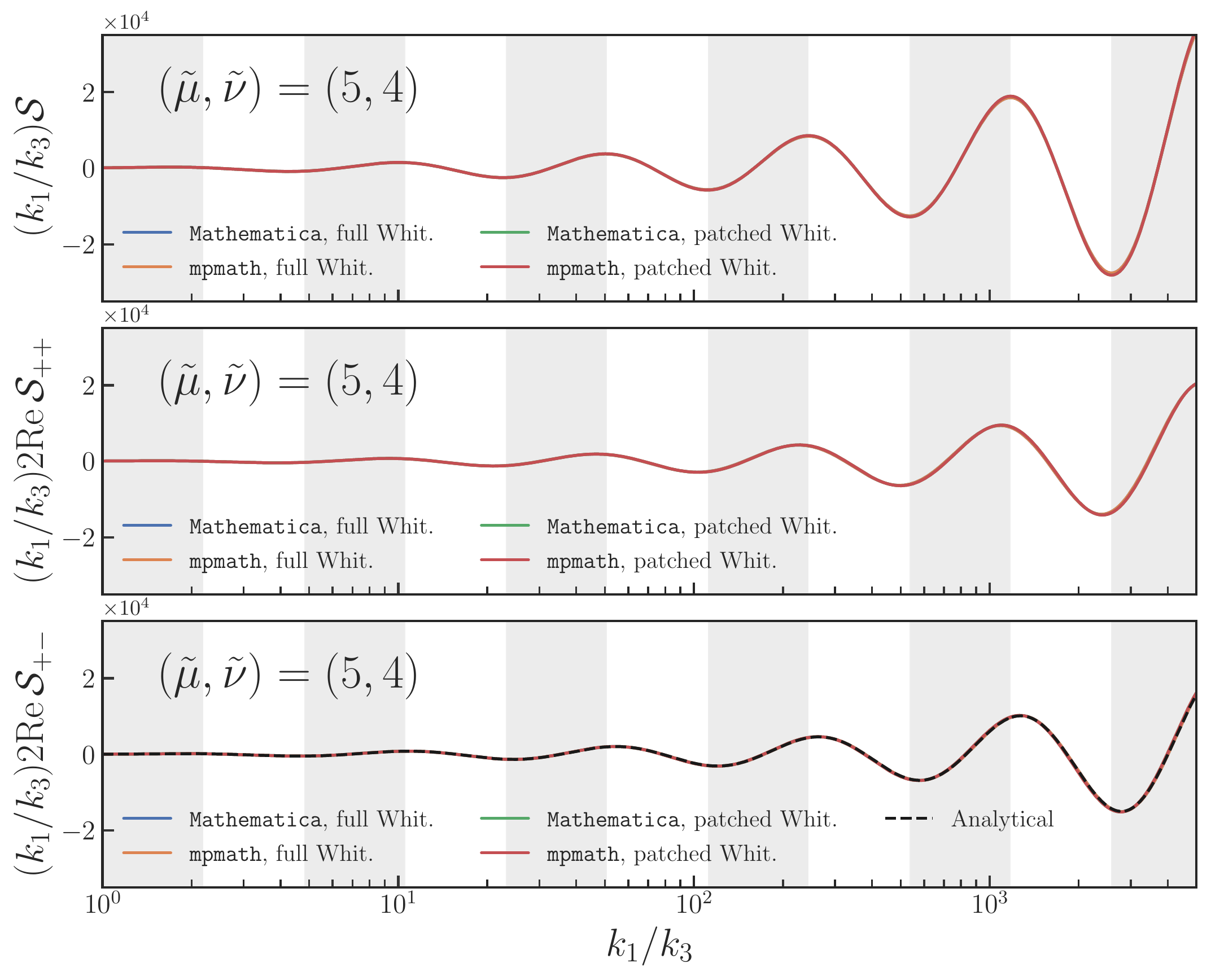} 
\caption{(\emph{Upper}) The CC signal $\mathcal S$, rescaled by $k_1/k_3$, in the squeezed limit, $k_1=k_2 \gg k_3$, for a tree-level diagram for complex scalar in an electric field. The electric field strength and the mass parameter are set such that $(\tilde \mu, \tilde \nu)=(5,4)$. We compare numerical results from \texttt{Mathematica~12} and those from \texttt{mpmath~1.1}, together with those from the full Whittaker functions and those from the patched Whittaker functions. (\emph{Middle}) $\sum \mathcal S_{\pm \pm}$ component of $\mathcal S$ ($\mathcal S = \sum \mathcal S_{\pm \pm} + \sum \mathcal S_{\pm \mp} =2 \text{Re} (\mathcal{S}_{++} +\mathcal{S}_{+-})$). (\emph{Lower}) $\sum \mathcal S_{\pm \mp}$  component of $\mathcal S$. We also include the analytical result from~\eqref{myany} (dashed black) to compare with the numerical results. A set of gray bands, which are evenly separated by $\Delta \log_{10} (k_1/k_3) = 2 \pi (\log_{10} e)/\tilde \nu$, are included for all panels as a guide to the oscillatory pattern.  }
\label{fig:mpmath}
\end{figure}

We employ \texttt{Mathematica} for most of the computational tasks through the paper. Here we crosscheck the numerical results from \texttt{Mathematica} code to those from  \texttt{mpmath} code. For each code implementation, we check results from computation with the full Whittaker functions as well as the patched Whittaker functions. The upper panel of~\figref{mpmath} shows the resulting signals for the parameter $(\tilde \mu, \tilde \nu)= (5, 4)$. The results from \texttt{Mathematica} are in good agreement with those from \texttt{mpmath} for both the full Whittaker and the patched Whittaker functions. The middle and lower panels of~\figref{mpmath} show such agreement exists at the individual SK diagram level. We again add the result from the analytical expression of $2 \text{Re}\,\mathcal S_{+-}$ to the lower panel of~\figref{mpmath}, which agrees well with the numerical results.

\section{The behavior of the loop integrand at large loop momentum}
\label{sec:large-mometum}
For the scalar or vector loop integral $\mathcal I^{(1)}_{\a\b}$, its integrand  can be expressed in terms of a single function $\mathcal{L}$ as
\begin{align}
\mathcal I^{(1)} (k_3; \tau_A, \tau_B) ={}& \int_{-1}^{+1} \di c_q \int_0^\infty \frac{\di q}{(2\pi)^2} \mathcal{L}(k_3; \tau_A ,\tau_B), \\
\mathcal{L}(k_3; \tau_A , \tau_B) \equiv{}&q^2 \tau_A^\alpha \tau_B^\alpha \frac{e^{2\tilde \mu \pi}}{{4p q}} \text{W}_{-\ii \tilde \mu, \ii \tilde \nu}(2 p \ii \tau_A)\text{W}_{\ii \tilde \mu, \ii \tilde \nu}(-2p\ii \tau_B)\text{W}_{-\ii \tilde \mu, \ii \tilde \nu}(2q\ii \tau_A)\text{W}_{\ii \tilde \mu,\ii \tilde \nu}(-2q\ii \tau_B),
\label{eq:largekey}
\end{align}
where $c_q \equiv \cos \theta_{\mb q}$, $p=\sqrt{q^2 - 2 q k_3 c_q +k_3^2}$, and the power $\alpha=2$ ($\alpha=0$) for the scalar (gauge boson) loop. The loop integral from the $(\a\b)=(++)$ diagram is given by
\beq
\mathcal I^{(1)}_{++} (k_3; \tau_1, \tau_2)=  \int_{-1}^{+1} \di c_q \int_0^\infty \frac{\di q}{(2\pi)^2}  \left[\left.\mathcal{L}(k_3; \tau_1, \tau_2)\right |_{\tau_1> \tau_2} +\left.\mathcal{L}(k_3; \tau_2, \tau_1)\right |_{\tau_2> \tau_1}\right],
\label{eq:ppsplit}
\eeq
Note that the splitting of the $\tau_1 > \tau_2$ and $\tau_2 > \tau_1$ parts of $\mathcal{L}$ is due to the $\theta$-functions inside the $D_{++}$ propagator. As discussed in Sec.~\ref{sec_setup}, we only need to perform the integral for the first (or the second) $\mathcal{L}$ term inside the integral (\ref{eq:ppsplit}) and the other part can be tabulated by copying. The loop integral from the $(\a\b)=(+-)$ diagram is given by
\beq
\mathcal I^{(1)}_{+-} (k_3; \tau_1, \tau_2) = \int_{-1}^{+1} \di c_q \int_0^\infty \frac{\di q}{(2\pi)^2}  \mathcal{L}(k_3; \tau_2, \tau_1).
\label{eq:pmsplit}
\eeq
Again we can split $\mathcal{L}$ into the $\tau_1 > \tau_2$ part and the  $\tau_1 < \tau_2$ part and only need to perform integral for one of them as discussed in Sec.~\ref{sec_setup}.

For our numerical procedure, it is important that the loop integral always yield finite results on a given time grid of $(\tau_1, \tau_2)$. Therefore, we need to check whether the loop integrand vanishes at the large loop momentum limit, $q\to \infty$. Under such limit, the other momentum inside the loop integral $p\approx q$ since $q-k_3\leq p \leq q+k_3$. (We fixed $k_3=1$.) Meanwhile, the Whittaker function can be expanded as $\text{W}_{\kappa,\lambda}(z)\approx z^\kappa e^{-z/2} \left[1+\mathcal O(z^{-1})\right]$ at $z\to \infty$. Under such limit, \eqref{largekey} becomes
\beq
\lim_{q\to \infty} \mathcal{L}(k_3; \tau_1, \tau_2) = \frac{\tau_1^\alpha \tau_2^\alpha}{4} \left(\frac{\tau_1}{\tau_2}\right)^{-2\ii \tilde \mu} e^{2\ii q (\tau_2-\tau_1)} +\mathcal O (q^{-1}).
\label{eq:large0}
\eeq

Let us first consider the large $q$ behavior of the integrand of $\mathcal I_{++}^{(1)}$.  The first $\mathcal{L}$ term inside the integral (\ref{eq:ppsplit}) is approximately given by
\begin{align}
\lim_{q\to \infty} \left.\mathcal{L}(k_3; \tau_1, \tau_2)\right |_{\tau_1> \tau_2} \approx {}& \left.\frac{\tau_1^\alpha \tau_2^\alpha}{4} \left(\frac{\tau_1}{\tau_2}\right)^{-2\ii \tilde \mu} e^{2\ii q (\tau_2-\tau_1)} \right|_{\tau_1 > \tau_2}.
\label{eq:large1}
\end{align}
Because the SK indices associated with $\tau_{1,2}$ are both $+$, we Wick-rotated $\tau_{1,2} \to -\ii \tau_{1,2}-\epsilon$ where  a small positive number  $\epsilon$  is added to avoid the branch cut of the Whittaker function at $z\in \mathbb R^-$. Under the Wick-rotation, ~\eqref{large1} yields
\beq
\lim_{q\to \infty} \left.\mathcal{L}(k_3; \tau_1, \tau_2)\right |_{\tau_1> \tau_2}  \approx \left.\frac{(-\tau_1\tau_2)^\alpha }{4} \left(\frac{\tau_1}{\tau_2}\right)^{-2 \ii \tilde \mu} e^{- 2 q (|\tau_2|-|\tau_1|)}\right |_{|\tau_2| > |\tau_1|} = 0
\eeq
given the suppression from $e^{-2 q (|\tau_2|-|\tau_1|)}$. For the same reasoning the second $\mathcal{L}$ term of the integral~(\ref{eq:ppsplit}), $\left.\mathcal{L}(k_3; \tau_2, \tau_1)\right |_{\tau_2> \tau_1}$, also vanishes at large $q$. Next, we consider the large $q$ behavior of the integrand of $\mathcal I_{+-}^{(1)}$. It is given by~\eqref{large0} up to the exchange of $\tau_1 \leftrightarrow \tau_2$. Given the SK indices  associated with $\tau_1$ and $\tau_2$ are $+$ and $-$ respectively, we Wick rotate $\tau_1 \to -\ii \tau_1$ and $\tau_2 \to +\ii \tau_2$ and have
\beq
\lim_{q\to \infty} \mathcal{L}(k_3; \tau_2, \tau_1)\approx \frac{(\tau_1\tau_2)^\alpha}{4} \left(-\frac{\tau_2}{\tau_1}\right)^{-2\ii \tilde \mu} e^{-2 q \left(|\tau_1| + |\tau_2|\right)} = 0
\eeq
given the suppression from $e^{-2 q (|\tau_1| + |\tau_2|)}$.

A tricky point comes from the integrand of $\mathcal I_{++}^{(1)}$ at the $\tau_1 \to \tau_2$ limit. Under such limit, the exponential  vanishes and no longer provides suppression as $q\to \infty$. The limit corresponds to the UV limit of the diagram, i.e., $\Delta \tau \to 0, q\to \infty$. As in quantum field theory in flat-space, we replace the propagators with the regulated propagators, under which the loop integrand becomes
\begin{align}
\mathcal{L}_\text{reg}(k_3; t_A ,t_B) ={}&q^2 \tau_A^\alpha \tau_B^\alpha \frac{e^{2\tilde \mu \pi}}{{4p q}} \left[\text{W}_{-\ii \tilde \mu, \ii \tilde \nu}(2 p \ii \tau_A)\text{W}_{\ii \tilde \mu, \ii \tilde \nu}(-2p\ii \tau_B)- \text{W}_{-\ii \tilde \mu, \ii \tilde \nu_\text{reg}}(2 p \ii \tau_A)\text{W}_{\ii \tilde \mu, \ii \tilde \nu_\text{reg}}(-2p\ii \tau_B)\right] \nonumber\\
&\left[\text{W}_{-\ii \tilde \mu, \ii \tilde \nu}(2q\ii \tau_A)\text{W}_{\ii \tilde \mu,\ii \tilde \nu}(-2q\ii \tau_B)- \text{W}_{-\ii \tilde \mu, \ii \tilde \nu_\text{reg}}(2q\ii \tau_A)\text{W}_{\ii \tilde \mu,\ii \tilde \nu_\text{reg}}(-2q\ii \tau_B)\right].
\label{eq:large3}
\end{align}
Expand the Whittaker function to $\text{W}_{\kappa,\lambda}(z)\approx z^\kappa e^{-z/2} \left[1+(\lambda^2-\kappa^2+\kappa-1/4)z^{-1} +\mathcal O(z^{-2})\right]$ under $z\to \infty$, \eqref{large3} at large $q$ is approximated by
\beq
\lim_{q\to \infty} \mathcal{L}_\text{reg}(k_3; \tau_1, \tau_2) \approx  \frac{\tau_1^\alpha \tau_2^\alpha}{4}\left(\frac{\tau_1}{\tau_2}\right)^{-2\ii \tilde \mu} e^{2\ii q (\tau_2 -\tau_1)} (\tilde \nu^2_\text{reg}-\tilde \nu^2)^2 \frac{[1-4\tilde \mu^2 +2\tilde \nu_\text{reg}^2 + 2 \tilde \nu^2+4\ii q (\tau_2-\tau_1)]^2}{64 q^4 \tau_1^2 \tau_2^2}.  
\eeq
The  additional $q^{-4}$ factor provides extra suppressions for the UV limit (and for integrands with $\tau_1\neq \tau_2$). Therefore the loop integrands always vanish at large $q$.

\section{Crosscheck the filtering methods}
\label{sec:crosscheckB}

In this section, we compare several signal filtering methods that separate the oscillatory part $O(x \equiv \log_{10} (k_1/k_3))$ from the non-oscillatory ``background" part $P(x)$ for a signal $S(x) \equiv (k_1/k_3) \mathcal{S} $ in the squeezed limit. Our first method is the {\bf derivative filtering}. If $|N|$ and $|L|$ of~(\ref{shapeofrho}) is much smaller than $\omega$, $P(x)$ is approximately a polynomial function of $x$ and vanishes after taking sufficient numbers of derivatives, while $S(x)$, in the form of $A \sin(\tilde \omega x+\varphi)$ with $\tilde \omega = \omega /\log_{10} e$, becomes $A \tilde \omega^n \sin (\tilde \omega x +\varphi')$ after taking $n$-th derivatives. The frequency of the oscillatory pattern $\tilde \omega$ keeps the same while its amplitude got enhanced if $\tilde \omega>1$. The derivates unavoidably introduce a phase shift, $\varphi \to \varphi'$, to the oscillatory pattern. To mitigate this complication, we take the fourth derivative of $S(x)$ and expect the oscillatory pattern to be recovered by  	
\begin{equation}	
\hat O(x) = \tilde \omega^{-4} \f{\partial^4}{\partial x^4} S(x).	
\label{eq:derivative_filter}	
\end{equation}

\begin{figure}[t!]
   \centering
\includegraphics[width=0.8\textwidth]{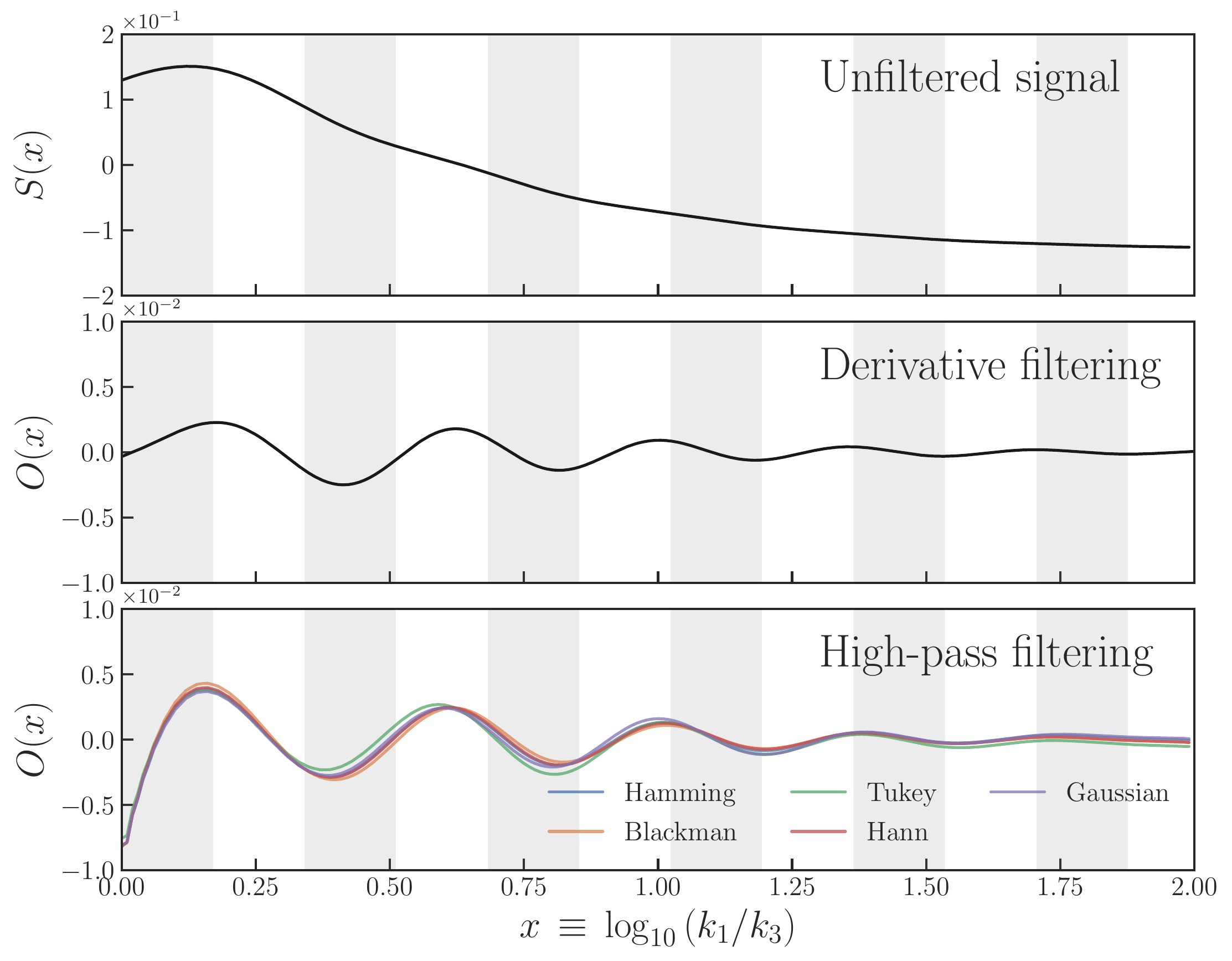} 
\caption{(\emph{Upper}) The rescaled CC signal $S = (k_1/k_3) \mathcal S (k_1, k_2, k_3)$ in the squeezed limit, $k_1=k_2 \gg k_3$, for the 1-loop diagram of gauge boson with parameter $(\tilde \mu, \tilde \nu, \tilde \nu_\text{reg})=(4,4,8)$. (\emph{Middle}) The oscillatory pattern after applying the derivative filtering~\eqref{derivative_filter} to the signal in the top panel. (\emph{Lower}) The oscillatory patterns after applying various high-pass filters of \texttt{Mathematica 12} to the signal in the top panel. The legend lists the names of the window functions used. A set of gray bands, which are evenly separated by $\Delta \log_{10} (k_1/k_3) = \pi (\log_{10} e)/\tilde \nu$, are included for all the panels as a guide to the oscillatory pattern.}
\label{fig:hpf}
\end{figure}

The upper panel of \figref{hpf} shows an unfiltered signal taken from the lower-left panel of~\figref{v_2}. It is for a pinched gauge boson 1-loop diagram with parameter $(\tilde \mu, \tilde \nu, \tilde \nu_\text{reg}) =(4,4,8)$ in the squeezed limit $k_1=k_2 \gg k_3$. The signal data is a discrete dataset with $x$ ranging from $x_{\min}=0$ to $x_{\max} =2$ with an even spacing of  $\Delta x = 0.01$. To apply the derivative filtering, we first interpolate the unfiltered signal. Next, we take the fourth derivatives of the interpolated function and rescale it with $\tilde \omega^{-4}$, where the oscillatory frequency $\tilde \omega = 2 \tilde \nu/\log_{10} e$. The result is shown in the middle panel of \figref{hpf}. In practice, one can extract $\tilde \omega$ by fitting $\f{\partial^4}{\partial x^4} S(x)$ and checked the pre-condition $|N|, |L| \ll \omega$ by fitting the resulting $S(x)-\hat O(x)$ and $\hat O(x)$.

An alternative method to access the oscillatory pattern without assuming particular functional forms for $P(x)$ and $O(x)$ is by using the {\bf high-pass filters}. A high-pass filter is a function defined in the frequency space of the signal. It has a cut-off frequency $\omega_c$ and attenuates signals below the cut-off frequency. The amount of attenuation depends on the filter window function. Usually, there is a trade-off between the attenuation and so-called ``ringing artifacts", which refer to the artificial oscillatory signals introduced by the filtering procedure. A classical example to illustrate this trade-off is the top-hat filter. The filter is a rectangle function in the frequency domain, $H(f) = \text{rect}(f)$, and a sinc function in the time domain, $h(t)=\text{sinc}(t)$. Convoluting the sinc function with a signal usually results in artificial oscillations. On the opposite end, a Gaussian filter, which is given by $G(f) = \exp \left(-\frac{f^2}{2\sigma_f^2}\right)$  in the frequency domain and $g(t)=\frac{1}{\sqrt{2\pi} \sigma} \exp\left(-\frac{t^2}{2 \sigma^2}\right)$ with $\sigma = \frac{1}{2\pi \sigma_f}$ in the time domain,  does not introduce any ringing artifacts but its power to attenuate low-frequency signal is also handicapped.

The lower panel of \figref{hpf} shows the results after passing the signal through a set of different high-pass filters that are built-in \texttt{Mathematica 12} (Hamming, Blackman, Tukey, Hann, and Gaussian)~\cite{hpf}. We set the filter kernel length to be the length of the signal dataset, and a cut-off frequency $\omega_c = 0.147$ for  function \texttt{HighpassFilter} for all filter choices. Note that  $\omega_c$ needs to satisfy $\omega_c \ll \tilde \omega /(2\pi)$ to avoid filtering out $O(x)$ together with $P(x)$. To be concrete, we choose $\omega_c = 0.05\, \tilde \omega /(2\pi) = 0.05 \tilde \nu/(\pi \log_{10} e)$ through the filtering procedures for the loop processes.	
The results of different high-pass filters are broadly consistent with each other. They are also in reasonable agreement with the result from the derivative filtering.  Therefore we choose the high-pass filter with a Gaussian window function through the filtering procedures in the main text. Note that for the simple high-pass filter we adopted, the resulting $O(x)$ may contain a small amount of the smooth component at small $x$ while the resulting $P(x) =S(x) - O(x)$ contains a small amount of the oscillatory pattern at large $x$. We leave a detailed study for better filtering procedures for the future.

\bibliography{cosmoco} 
\bibliographystyle{utphys}

\end{document}